\documentclass[journal]{IEEEtran}

\usepackage{graphicx, caption, subcaption}
\usepackage{algorithm}
\usepackage{array}

\usepackage{algorithmicx}
\usepackage{algpseudocode}
\usepackage[utf8]{inputenc}
\usepackage{soul}
\usepackage{verbatim}
\usepackage{cite}
\usepackage{listings}
\usepackage[table]{xcolor}
\usepackage[nolist]{acronym}
\usepackage{amsmath}
\usepackage{amssymb}
\usepackage{soul}
\usepackage{rotating}
\usepackage{multirow} 
\usepackage{comment}
\usepackage{listings}
\usepackage{xcolor}

\definecolor{background}{RGB}{39, 40, 34}
\definecolor{string}{RGB}{230, 219, 116}
\definecolor{comment}{RGB}{117, 113, 94}
\definecolor{normal}{RGB}{248, 248, 242}
\definecolor{identifier}{RGB}{166, 226, 46}
\definecolor{ascolor}{RGB}{230, 219, 116}     

\lstdefinelanguage{PythonCustom}{
    language=Python,
    morekeywords={as}, 
    keywordstyle=[2]\color{ascolor}\bfseries, 
    keywords=[2]{as} 
}

\usepackage{mathtools, stmaryrd}
\usepackage{xparse} \DeclarePairedDelimiterX{\Iintv}[1]{\llbracket}{\rrbracket}{\iintvargs{#1}}
\NewDocumentCommand{\iintvargs}{>{\SplitArgument{1}{,}}m}
{\iintvargsaux#1} %
\NewDocumentCommand{\iintvargsaux}{mm} {#1\mkern1.5mu..\mkern1.5mu#2}

\usepackage{siunitx}

\usepackage{blindtext}
\usepackage{ctable} 

\def\BibTeX{{\rm B\kern-.05em{\sc i\kern-.025em b}\kern-.08em
    T\kern-.1667em\lower.7ex\hbox{E}\kern-.125emX}}

\newcommand\Tstrut{\rule{0pt}{2.6ex}}         
\newcommand\Bstrut{\rule[-0.9ex]{0pt}{0pt}}   

\begin{document}

\title{Optimizing LoRa for Edge Computing with TinyML Pipeline for Channel Hopping  \\
}

\author{\IEEEauthorblockN{%
  Marla Grunewald,
  Mounir Bensalem 
   and   Admela Jukan}\\
\IEEEauthorblockA{%
   Technical University of Braunschweig, Germany\\ \{marla.grunewald, mounir.bensalem,  a.jukan\}@tu-bs.de} %
}

\maketitle
\thispagestyle{empty}
\begin{abstract} We propose to integrate long-distance LongRange (LoRa) communication solution for sending the data from IoT to the edge computing system, by taking advantage of its unlicensed nature and the potential for open source implementations that are common in edge computing. We propose a channel hoping optimization model and apply TinyML-based channel hoping model based for LoRa transmissions, as well as experimentally study a fast predictive algorithm to find free channels between edge and IoT devices.  In the open source experimental setup that includes LoRa, TinyML and IoT-edge-cloud continuum, we integrate a novel application workflow and cloud-friendly protocol solutions in a case study of plant recommender application that combines concepts of microfarming and urban computing. In a LoRa-optimized edge computing setup, we engineer the application workflow, and apply collaborative filtering and various machine learning algorithms on application data collected to identify and recommend the planting schedule for a specific microfarm in an urban area. In the LoRa experiments,  we measure the occurrence of packet loss, RSSI, and SNR, using a random  channel hoping scheme to compare with our proposed  TinyML method.   The results show that it is feasible to use TinyML in microcontrollers for channel hopping, while  proving the effectiveness   of TinyML in learning  to predict the best  channel to select for LoRa transmission, and by improving the RSSI by up to 63 \%, SNR by  up to 44 \% in comparison with a random hopping mechanism.
 \end{abstract}
\section{Introduction} \label{sec:intro}
Lora is a well-known long-distance communication protocol that relies on unlicensed spectrum and CSS modulation to facilitate communication over longer distances\cite{Toong2024LoraPerformance}. In the rapidly evolving field of IoT-edge-cloud computing, also known as the compute continuum, the potential for integrating LoRa with IoT, edge and cloud computing carries the promise of further expanding the possibilities for new applications to evolve \cite{Sarker2019survey, Jindal2022edgelora}. When combined with edge computing, LoRa can help more efficient bandwidth utilization, whereby some tasks, like those that are machine learning based, can be offloaded from cloud to the edge. Especially in the context of compute contiuum which has enabled a wide availability of the open source libraries, integration of LoRa in the context of edge computing could also become more broadly available through open source. Also edge computing can benefit from LoRa communications by improving its connectivity and server load balancing, especially in remote areas.  

Optimizing LoRa for edge computing introduces multiple new research challenges.  Currently, there are no protocols that convert LoRa communications into something usable by edge computing and such adaptations need to be done manually. Also, edge devices need an antenna, i.e., additional hardware, to operating at LoRa frequency. There are duty cycle restrictions on almost all frequency ranges due to its unlicensed nature, which makes the communication fundamentally unreliable and unscheduled. LoRa devices need to time synchronize, which makes the management of sending/receiving between the edge and IoT devices a challenge.  From a more practical perspective, the availability of open source libraries as well as the hardware support are critical. For applications, which increasingly integrate ML and AI tools in a coordinated cloud-edge continuum, the data collection in IoT devices requires a more effective processing in the edge, which can also benefit from a collision-free LoRa transmission, especially when scaling up the number of devices. 
\par In this paper, we optimize the performance of long-distance LongRange (LoRa) communication solution for sending the data from IoT to the edge computing system for a cloud-based application developed for urban living. The IoT devices include sensors that collect and send data to the edge devices, for data processing, prediction and interface. We implement a machine learning algorithm based on TinyML for collision-free channel assignment for LoRa transmissions to the edge, and propose a predictive mechanism to this end.  In an open source experimental setup that includes LoRa, TinyML and IoT-edge-cloud continuum implementation, we integrate a novel plant recommendation application for urban gardening. We refer to our system as an urban computing continuum, which we define as a data, communication and computer infrastructure for urban microfarming. In our urban computing continuum, we propose and analyze a novel system architecture that can efficiently and seamlessly integrate LoRa communication, edge and cloud computing in one open system. The results show improvements transmission performance and reliability and a comparison is done do a random channel hopping mechanism. Furthermore, we show the performance of cosine similarity in predicting plant performance in different soils with different data sparsity values and compare a collection of ML algorithms for usability in the proposed continuum.
The rest of the paper is organized as follows. Section \ref{sec:relwork} presents related work. Section \ref{sec:sys} introduces the reference system architecture, including channel hopping optimization, the proposed TinyML pipeline for channel hopping and a discussion on the TinyMl placement in the system. Section \ref{sec:ml} presents a case study demonstrating the user of LoRa to transmit soil data for integration int an urban computing continuum, with the objective of recommending suitable plants to a user. The results of the experimental and simulated studies conducted on the proposed TinyML model and the case studies are presented in Section \ref{sec:results}. Section \ref{sec:conclusion} concludes the paper.

\section{Related Work} \label{sec:relwork} 
The integration of LoRa with IoT, edge and cloud computing continuum is not a new endeavor, but has not been largely investigated in the research community. In \cite{Sarker2019survey}, it was found that edge computing can facilitate more efficient bandwidth utilization in LoRa by offloading certain computational tasks from the cloud to the edge. Paper \cite{Jindal2022edgelora} employs edge computing with LoRa to transmit low resolution images, thereby reducing the overall energy consumption by performing image recognition operations at the edge rather than in the remote cloud. Lack of practical integration of LoRa can also be partially attributed to the already mentioned unavailability of the open source libraries, which however are common place in edge computing. In LoRa, the communication process involves a sending node, a LoRa gateway, and in the case of LoRaWAN, a server that is responsible for data processing and acts as a user interface \cite{Santoro2023NWsimulator}. Again, the number of available open source LoRa gateways remains limited \cite{loramqtt}. Furthermore, the support for the open source gateways is not always guaranteed, such as Chirpstack; as of November 2024, Chirpstack provides support for a certain hardware category, such as LoRa shields from, e.g., Dragino or RAK wireless for a Raspberry Pi. 
In this paper, we use our open source implementation of the open-source LoRa gateway, as detailed in paper \cite{loramqtt}, to receive LoRa packets and forward them to the edge, where they are subsequently processed. 

\par To optimize LoRa performance for edge computing, our work uses frequency hopping for faster switching the carrier frequency across multiple frequencies within a wide spectral band. One of the key decisions to be made in this regard is how to frequency hop and the subsequent execution are conducted in the edge node. Our approach to this end is in application of machine learning with TinyML for the purpose of channel assignment for forthcoming transmissions to ensure reliable communication and reduce collisions among LoRa transmissions. Our predictive mechanism which aims at minimizing collision is different from previous work, such as in \cite{Farooqchannelaccess2018} where a LoRa node randomly selects a frequency, or \cite{AhmarFreHopping2019}  which implements a cryptographic frequency hopping solution for the EU863-870MHz frequency spectrum. Paper \cite{MhatreFreHopping2023} proposes a reinforcement learning-based algorithm that employs frequency hopping with consideration of   factors such as bandwidth, frequency, transmission power and time slots to predict the optimal and collision-free channel for allocation;  the reward function is based on energy consumption. While our objective shares a resemblance to this work, there are notable differences in our methodology and the factors we consider. In our approach, we employ a hopping algorithm with the objective of minimizing collision. This is achieved through the integration of channel sensing techniques, implemented at the edge and IoT nodes. In making the hopping decision, we take into account the base frequency as well as the history of channel occupation, which is novel. 

Tiny Machine Learning (TinyML)  is an increasingly prominent area within machine learning technologies and applications \cite{abadade2023comprehensive}, dedicated to enabling efficient data analysis on ultra-low-power devices. It integrates specialized HW, compressed algorithms, and tailored software to process data from always-on sensors, such as vision, audio, and biometric sensors, while maintaining power consumption in the milliwatt range or lower. Thus, TinyML is a suitable solution to adopt for leveraging intelligence in the end-nodes, and particularly for channel hopping decision making for LoRa networks. 
Paper \cite{lodhi2024tiny} used tinyML for optimizing channel allocation within LoraWAN scenario, where end-nodes uses tiny ML to select LoraWAN channels, using ns3 simulations. In our work, as we implement the physical layer Lora for transmitting sensor data, using an experimental setup, considering sensing channel occupation state, and in order to decrease the number of collisions, we also propose a tiny ML based channel hopping solution, which we also experimentally evaluate. The placement of channel allocation function is also an open issue, i.e., whether in the end-node or gateway. Paper \cite{aihara2019q} developed a Q-learning-based channel allocation mechanism in the gateway (GW) that improves packet delivery ratio (PDR) performance by using the number of successfully received packets as the reward signal. In our approach, we analyze the optimal placement for the  channel allocation mechanism, and provide a solution that shows that placing channel allocation function at the EN with the help of TinyML is better to providing a scalable and easy-to-integrate solution.
\par As proof-of-concept, we use an application case study inspired by the concepts of urban computing and microfarming, which we proposed in our previous paper \cite{latincom}. In  \cite{latincom} our main motivation was the application performance of the urban \emph{pendant} to farming referred to as microfarming, a sustainable farming concept at a smaller scale implemented in urban gardens, typically at less than 5 acres of garden size \cite{jorda2019automated, gkonis2023survey}.  We combined this application with the concept of urban living \cite{su15053916}, which ultimately led to the concept of urban computing continuum which we coined. In this paper, we majorly extend our previous paper  \cite{latincom} to focus on LoRa optimizations for this class of applications. More in detail, we focus on maximizing the channel occupancy using TinyML based frequency hopping. Furthermore, the communication and computation between the edge and the Lora gateway are integrated. 
Furthermore, the cloud-based functions of the application are extended through the deployment of cosine similarity on an extended dataset, evaluating five levels of sparsity in our data.
The database of plants which we use here has also been majorly extended and used to train the application ML models on this new data set for plant recommendations. All experimental setup and related code are open source 
\footnote{https://gitlab.com/M4RL4/loraedgetinyml}.

\section{System Description}\label{sec:sys}

\begin{figure}
    \centering
    \includegraphics[scale=1.1]{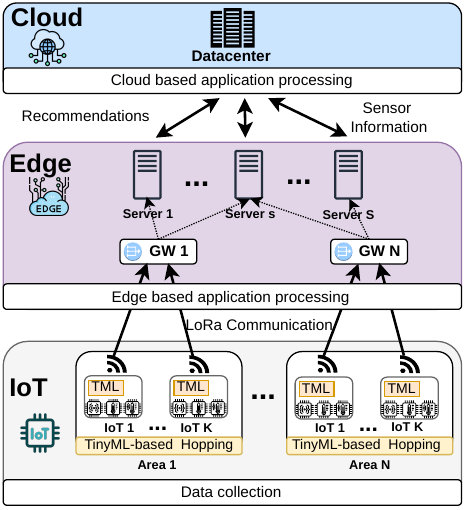}
    \caption{LoRa for Edge Computing in a Compute Continuum.}
    \label{fig:architecture}
\end{figure}

\subsection{Reference Architecture}
Figure \ref{fig:architecture} illustrates the integration of LoRa in an IoT/edge/cloud compute continuum today. The continuum system architecture generally encompasses three types of devices with a range of capabilities that can jointly process, store and communicate data: IoT,  edge, and cloud. Each type of devices has a specific computing capability and constraints as well as different delay, jitter and packet loss along the connections. IoT devices are constrained devices that collect data and send them to the more powerful type of devices, be in the edge or the cloud, such as for storage and processing.    We assume that IoT devices are spread in $N$ geographic IoT areas, where in each area there is a heterogeneous set of $K$ devices, each equipped with a sensor and microcontroller unit. These units are collecting and sending  data on a periodic basis  to an edge server.  The edge server can be located in close proximity to the IoT areas, such as in communal buildings in an urban setting. Let us assume that each IoT area has its own server, resulting in a total of $S$ edge servers for $N$ areas. Communication between the edge and the IoT devices  is assumed using LoRa. For every area $N$, LoRa implements one gateway, $GW$ in the edge. This gateways are hardware and software implementation of the LoRa transciever and can be implemented per server, or a group of servers in edge computing.  In our approach, we implement one gateway per edge server. From the application perspective, data is collected in the IoT areas and pre-precessed in the edge computing part of the continuum, such as for machine learning inference to preprocess data and retrieve insight in the decision making process before moving data from the cloud. Finally, cloud based application processing enable more compute intensive tasks, such as training for machine learning. 

To optimize LoRa for edge computing, the fundamental aspect to consider is the reliable transmission between edge and IoT devices, which requires the selection of optimal channel characteristics. The parameters that can be modified include bandwidth, coding rate, spreading factor and base frequency. Previous studies have demonstrated that certain combination of these parameters can impact the quality and range of LoRa transmissions. For instance, a spreading factor of 12 has been shown to enhance range at the cost of a lower data rate, whereas a spreading factor of 7 has been observed to reduce it therefore providing a bigger data rate\cite{Selim2024SF}. While the identification of the optimal range and data rate combination for a single LoRa transmission in a remote area may be a relatively straightforward process, the same cannot be said for densely populated areas, where the task becomes considerably more challenging. LoRa operates on the license-free spectrum, which consequently results in competition with other radio modules that share the same frequency band. Consequently, the deployment of a larger-scale LoRa network in a densely used frequency band may encounter distinct challenges. To achieve a flexible and efficient LoRa network capable of operating in a densely used area, we propose a new frequency hopping mechanism that continuously seeks to identify less utilized sub-frequencies within a specified frequency band. This approach aims to balance the utilization of existing frequency bands, minimizing collisions and to ensure a fair and reliable usage of the given frequency band, by assuring stability, i.e., minimizing channel hopping. In addition to the analytical approach, this section also describes the channel hoping mechanisms that use TinyML (TML) implementation at the IoT nodes, and later analyzes the quality of the results obtained. The notation used is shown in Table \ref{tab:notations}. 

\subsection{Channel Hopping Optimization}
\subsubsection{Problem Formulation}
We define a set of LoRa end-nodes (EN)s in the IoT context, $\mathcal{E}=\{EN_{1}, ..., EN_{N}\}$, and a set of LoRa gateways (GW)s in the edge computing context, $\mathcal{G}=\{GW_{1}, ..., GW_{G}\}$. We define a set of frequencies to be used by end-nodes for LoRa transmission as $\mathcal{F}=\{f_1,...,f_F \}$, e.g., \{868.1, 868.3, 868.5\} MHz.  
We define a binary decision variable $x_{i,g, f}(\tau)$ for the channel allocation problem, given by:  
\begin{equation}
    x_{i,g, f}(\tau) = \begin{cases}
1 &\text{if EN$_i$  sends  to gateway GW$_g$} \\
 &  \text{ using frequency $f$ at time slot $\tau$}\\
0 &\text{otherwise}
\end{cases}
\end{equation}
 A collision occurs when two transmission events overlap, which is equivalent to the case when two ENs sends data to the same gateway at the same time slot $\tau$ using the same frequency. We denote by $c_{i,j,g,f}(\tau) $ the collision factor between two ENs $i$ and $j$, sending to the gateway GW$_g$ at time slot $\tau$, which is given as follows:
\begin{equation}
    c_{i,j,g,f}(\tau) = x_{i,g,f}(\tau) x_{j,g,f}(\tau)
\end{equation}
The first objective of the channel collision avoidance problem is to minimize the number of collisions that occur  when two IoT end-nodes are sending to the same edge gateway at the same time using the same frequency, during a period of time, and given by: 
\begin{equation}
  \min_{} \left( \mathcal{O}_{1}= \sum_{\tau=1}^{T} \sum_{i=1}^{N}\sum_{j=1}^{N}\sum_{g=1}^{G}\sum_{f=1}^{F} c_{i,j,g,f}(\tau) \right)
\end{equation}
Channel hopping can be used as a solution to avoid collision, whenever an end-node is sensing that the channel is already in use by another node.  We introduce a hopping indicator binary variable $z_{i}(\tau)$ that gives 1 if the EN$_i$ hops to another channel at time slot $\tau$. Thus, the second objective function is the  minimization of number of channel hoppings, assuming that the action of hopping is penalized as it introduces an overhead to the system. 
\begin{equation}
  \min_{} \left( \mathcal{O}_{2}=  \sum_{i=1}^{N}\sum_{\tau=2}^{T}  z_{i}(\tau) \right)
\end{equation}


We assume that each EN can use only one frequency to transmit the signal to one or more gateways, which can be given by the following constraint.
\begin{equation}\label{eq:sending}
    \sum_{f=1}^{F} x_{i,g, f}(\tau) =1, \forall \text{EN}_i\in \mathcal{E}, \forall \text{GW}_g \in \mathcal{G}, \forall \tau
\end{equation}
We assume that a gateway $g$ has limited channel capacity, represented by $M_g$ the number of channels at GW$_g$. Thus, we obtain the following constraint, considering the channel allocation a binary variable.
\begin{equation}\label{eq:gw}
    \sum_{i=1}^{N}\sum_{f=1}^{F} x_{i,g, f}(\tau) \leq M_g,  \forall \text{GW}_g \in \mathcal{G}, \forall \tau
\end{equation}
We define an integer decision variable $s_{i,g, f}(\tau)$ to set the number of symbols sent by EN$_i$ to gateway GW$_g$  using frequency $f$ at time slot $\tau$. The data transmitted to each gateway can be constrained by certain value that is frequency dependent, and can be expressed as follows:
\begin{equation}\label{eq:gwbf}
   \sum_{i=1}^{N} s_{i,g, f}(\tau) \leq B_f  ,  \forall \text{GW}_g \in \mathcal{G}, \forall \tau \forall f
\end{equation}
From the IoT side, the data transmission capacity constraint per channel is given by:
\begin{equation}\label{eq:gwbmin}
  B_{\text{min}} \leq s_{i,g, f}(\tau) \leq B_f, \forall \text{EN}_i\in \mathcal{E}, \forall \text{GW}_g \in \mathcal{G}, \forall \tau \forall f
\end{equation}
where $B_{\text{min}}$ is the minimum number of symbols per packet. \\ 
Let us assuming that each end-node has a certain amount of data $D_i$ to transmit during a period of time. We define the following constraint to achieve the full data transmission $D_i$. 
\begin{equation}\label{eq:data}
   \sum_{\tau=1}^{T} \sum_{g=1}^{G} \sum_{f=1}^{F} x_{i,g, f}(\tau) s_{i,g, f}(\tau) = D_i, \forall \text{EN}_i\in \mathcal{E}
\end{equation}

In this model, we consider that a channel hopping decision is taken when a collision occurs on a channel from the IoT end-node towards a LoRa gateway in the edge.  If a collision happens on a frequency channel $f$, at time of sensing $\tau-1$ for any gateway $g$, i.e., $\sum_{i=1}^{N} x_{i,g, f}(\tau-1)\geq 2$, at time $\tau$, we force all ENs to frequency hop except for one EN; the latter executes a sensing-based channel hopping algorithm. To model this, we define an extra binary variable $\delta_{g,f}(\tau) \in \{0,1\}$ to  force the model to hop the channels that are affected by collision. The three equation in (\ref{eq:hopping}) are  the linear programming formulation of the statement \textit{"if a$>$b then c=d"}, where $a=\sum_{i=1}^{N} x_{i,g, f}(\tau-1)$ is the number of nodes using the same channel at the same time, $b=2$ meaning that if more than 2 nodes are using the same channel at the same time then the number of collisions is bigger than 1, $c= \sum_{i=1}^{N} x_{i,g, f}(\tau)$ is the number of nodes using the same channel at the same time at time slot $\tau$, and $d=1$ meaning that we force the system to leave only one node connected through the same channel, which is equivalent to hopping all the colliding channels. 

\begin{equation}\begin{split}
  \sum_{i=1}^{N} x_{i,g, f}(\tau-1) & \geq 2- M (1-\delta_{g,f}(\tau) )\\
  \sum_{i=1}^{N} x_{i,g, f}(\tau-1) &  \leq 2 + M \delta_{g,f}(\tau) \\
  1 - M(1-\delta_{g,f}(\tau) ) & \leq \sum_{i=1}^{N} x_{i,g, f}(\tau)  \leq 1 + M(1-\delta_{g,f}(\tau) )
  \end{split}\label{eq:hopping}
\end{equation}
where $M$ is a big number.

In order to account the number of channel hoppings during a period of time, we  define the constraints that set the value $z_{i} (\tau)$. If a channel is hopped, there is at least $ x_{i,g, f}(\tau)-x_{i,g, f}(\tau-1)=1$, so that $z_{i} (\tau)$ is bigger or equal to 1, and the expression is given as follows:
\begin{equation}\label{eq:hopcount}
    x_{i,g, f}(\tau)-x_{i,g, f}(\tau-1) \leq z_{i} (\tau) , \forall \text{EN}_i\in \mathcal{E} 
\end{equation}
If a channel is not hopped, all $ x_{i,g, f}(\tau)-x_{i,g, f}(\tau-1)$ will be 0 so $z_{i} (\tau)$  will be less or equal to  0, otherwise at least one channel will hop and with the absolute value, $z_{i} (\tau)$ bounded by at least  1. 

\begin{equation}\label{eq:hopcount1}
    z_{i} (\tau)\leq \sum_{g=1}^{G}  \sum_{f=1}^{F}   \left| x_{i,g, f}(\tau)-x_{i,g, f}(\tau-1)  \right|, \forall \text{EN}_i\in \mathcal{E} 
\end{equation}

Finally, the Channel Hopping Optimization Problem can then  formulated as follows:
\begin{equation}\begin{split}
    \min_{x, s } & \left(  \alpha \cdot  \mathcal{O}_1 + \beta \cdot \mathcal{O}_2  \right)  \\
    \text{subject to:   }& {Eq.} (\ref{eq:sending}), (\ref{eq:gw}), (\ref{eq:gwbf}),(\ref{eq:gwbmin}), (\ref{eq:data}), (\ref{eq:hopping}), (\ref{eq:hopcount}), (\ref{eq:hopcount1})
\end{split}
\end{equation}
where $\alpha $ and $\beta$ are weights associated to the two objective functions.  The inequality (\ref{eq:gw}) and  (\ref{eq:gwbf}) represents capacity constraints inequalities which are similar to Knapsack inequality, allowing us to polynomialy  reduce the problem to a Knapsack problem if we relax all the other constraints, which is still an NP-Complete problem. A MILP solution would be computationally high, and cannot run on IoT devices. Furthermore, a MILP solution would not be scalable if we increase the number of channels to be used,  number of end-nodes, number of gateways, or the number of time slots. We therefore resort to a more practical, ML-based solution to tackle the channel hopping decision making problem, as explained next. 


\begin{table}[h!]
  \begin{center}
    \caption{Notations}
    \label{tab:notations}
    \begin{tabular}{c|l}
    \hline 
      \textbf{Notation} & \textbf{Description} \Tstrut\Bstrut \\ \hline
      \Tstrut\Bstrut
      & \multicolumn{1}{c}{\textbf{\underline{Sets}}}\\ 
     $\mathcal{E}$  &   A set of end-nodes (EN)s   \Tstrut \\
      $\mathcal{G}$ &  A set of gateways (GW)s\\
      $\mathcal{F}$ &  A set of frequencies\\ 
      \hline
      \Tstrut\Bstrut
       & \multicolumn{1}{c}{\textbf{\underline{Parameters}}}\\ 
     $N$  &   Number of end-nodes   \Tstrut \\ 
     $G$  &   Number of gateways   \\ 
      $F$  &   Number of  frequencies \\
      $B_f$  &  The maximum number of symbols supported \\&by a frequency $f$ per time unit.  \\
      $B_{\text{min}}$  &   The minimum number of symbols per packet \\
      $D_i$  & The amount of data  to transmit from EN$_i$ \\ & during a period of time. \\ \hline
      \Tstrut\Bstrut 
      & \multicolumn{1}{c}{\textbf{\underline{Variables}}}\\ 
     $x_{ifg}(\tau)$  &  A binary variable for channel allocation of sender EN$_i$ \Tstrut \\ &  to GW$_g$ through frequency $f$ at time slot $\tau$ \\
     $ s_{i,g, f}(\tau)$ & An integer decision variable that define the number of  \\ & symbols sent by EN$_i$ to gateway GW$_g$  using frequency $f$  \\ & at time slot $\tau$  \\
     $z_{i} (\tau)$ & A binary variable for channel hopping for EN$_i$ at time slot $\tau$\\
       $\delta_{g,f}(\tau)$ & An extra binary variable to help modeling channel hopping  \Bstrut \\
      \hline
    \end{tabular}
  \end{center}
\end{table}


\subsection{Tiny ML Pipeline for Channel Hopping}
\begin{figure}
    \centering
    \includegraphics[width=\linewidth]{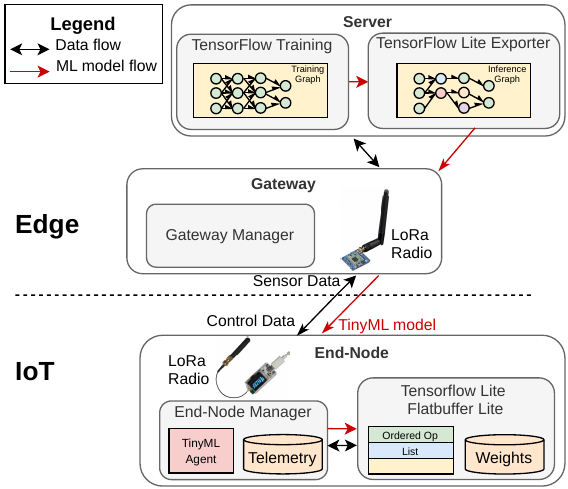}
    \caption{TinyML flowchart}
    \label{fig:TMLworkflow}
\end{figure}

We propose to automate the channel hopping decision using TinyML. The TinyML pipeline is illustrated in Figure \ref{fig:TMLworkflow}. As the ML training phase requires a certain computing requirements that cannot always be met by IoT end nodes, such as micro-controllers ESP32,  and because the energy consumption is critical at the IoT layer,  we chose to place the training and compression of the channel hoping ML models in a server on the edge layer, and place the TinyML model in the end-nodes. The channel hopping based on TinyML inco through three main components: data collection and cleaning, training and compression and in-chip inference. 

\subsubsection{Data Collection and Cleaning} 
The input data for channel hopping algorithm consists of three types of information: the time series historical data of channel availability, the historical data of SNR and RSSI. The channel availability information of frequency $f$ received by an end-node E$_i$ from all gateways at time slot $\tau$ is modeled as: 
\begin{equation}\begin{split}
\Delta_{i,f}(\tau)=& \sum_{j=1}^{N}\sum_{g=1}^{G} x_{j,g,f}(\tau)
\end{split} 
\end{equation}
Each end-node collects the channel availability data from different time slots. We assume that, at time $t$, we consider all data collected from $ts$ time slots to be used in channel hopping, due to the assumed limited storage capability of the IoT nodes. The channel availability of all frequencies of a single data point and of the considered duration are represented as follows:
\begin{equation}\begin{split}
\Delta_{i}(\tau)= & \{ \Delta_{i,f}(\tau)\}_{f\in [1..F]}, \\
     \hat{\Delta}_{i}(t)= & \{ \Delta_{i}(t-ts \cdot \tau),..., \Delta_{i}(t-k\tau),...,\Delta_{i}(t)\}  
\end{split} 
\end{equation}
The two other important type of data that will be used for channel hoppping decision are the RSSI and SNR values, which can be measured at the gateway and sent to the end-nodes as control messages. The RSSI and SNR will be stored in the end-nodes and updated periodically so that we keep a fixed amount of data entries, and can be presented as follows:
\begin{equation}
    \text{RSSI}_i(t)=\{\text{RSSI}_i(t-ts \cdot \tau),...,\text{RSSI}_i(t) ) \}
\end{equation}
\begin{equation}
    \text{SNR}_i(t)=\{\text{SNR}_i(t-ts \cdot \tau),...,\text{SNR}_i(t)) \}
\end{equation}
After collecting all the required data, a data set $\mathcal{S}$ is obtained. The data set $\mathcal{S}$ represented as follows will be   formatted as a JSON file and used as an input for TinyML. 
\begin{equation}\label{eq:state}
    \mathcal{S}=\{s_i=(  \hat{\Delta}_{i}(t),  \text{RSSI}_i(t),  \text{SNR}_i(t)), \forall E_i,  \forall t \}
\end{equation}
The output of the channel hopping scheme  represents the channel decision  $\mathcal{C}_i$ at time $t$ for EN$_i$. In order to collect  the data of  best channel choice at every time slot and every network state, we repeat the scenario with all possible options i.e. all possible channels, and then we label the data point with the channel that has highest RSSI. The final channel vector associated to our data set is given as follows:
\begin{equation}
    \mathcal{C}=\{  \mathcal{C}_i= c_{i}(t), \forall E_i,  \forall t \}
\end{equation}

\subsubsection{Training and Compression} 
As developing a robust model converter requires significant engineering effort,  we adopt the existing TensorFlow Lite (TFLite) toolchain \cite{david2021tensorflow}.  The TFLite toolchain facilitates model compression and optimization, producing a FlatBuffer file that (TensorFlow Lite Micro) TFLM uses to load inference models in a micro-controller, such as ESP32 in our case. TensorFlow Lite enables a unified environment for both model development and execution. We adopt a fully connected neural network architecture (FCNN) modeled using tensorflow and TFLite, and trained on the data set $\mathcal{S}$ as input and $\mathcal{C}$ as output, where the output layer is represented by a multi-class layer to model all possible channel selection possibilities. As the training is going to be in a capable server in the edge, a Tensorflow Lite Exporter is used to create a small version of the model with a size conforming to the micro-controller capacity and LoRa packet limitations. The created model will be then updated in the micro-controller through  over-the-air (OTA) updates feature, by sending it through the gateway LoRa radio to the end-nodes.

\subsubsection{In-Chip Inference }
As all gateways send control data as shown in figure \ref{fig:TMLworkflow},  we propose a module called  End-Node Manager to control the communication between the EN and the gateways, by storing control logs and running channel hopping algorithm with the help of TinyML. The End-Node Manager consists of two parts: a TinyML agent and a Telemetry Database. The tinyML agent is responsible on executing the channel hopping mechanism with the help of the trained TFLite model, which will decide the next channel to hop.  The Telemetry Database is a small database designed to store the needed information described in (\ref{eq:state}), representing the input parameters for the TinyML model. The number of time steps $ts$ is constrained by memory size of the micro-controller which defines the temporal memory used for prediction. Each time the Gateway sends a sensing data (control data) containing the channel availability, RSSI, SNR of the previous transmission, the TinyML agent will store them in the Telemetry and deletes the oldest entries to keep the size at the defined level. 

\subsection{Discussion on TinyML Placement}

\begin{figure}
    \centering
    \includegraphics[width=\linewidth]{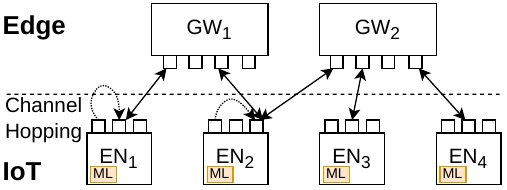}
    \caption{Placement of TinyML in IoT LoRa end-nodes }
    \label{fig:TML1}
\end{figure}

As previously mentioned, we place TinyML model in IoT end-nodes, as illustrated in Figures \ref {fig:TMLworkflow} and generalized in \ref{fig:TML1}. However, also the placement  at LoRa Gateways is an option. We now briefly discuss pros and cons of both choices.  

The primary benefit of frequency hopping at the IoT end node is that if facilitates scalability, as all end nodes can be flashed using a unified code, which has practical significance. 
Furthermore, this makes it possible that the communication between the devices may adapt to the environmental conditions over time with the objective of achieving a balanced distribution of frequencies within a frequency band, even in comparison with other technologies operating in the same band. This approach could facilitate not only a reliable, low-collision operation of our communications, but also the avoidance of interference with other systems.
This approach also facilitates scalable integration of newly added nodes. As a result of the bi-directional communication, the gateway is capable of send messages to the end-node about channels that are less frequently used. Consequently, the edge (gateway) and IoT (end-node) can transition to a less congested frequency when required, through the gateways monitoring of available frequencies and scanning of the spectrum.

Placing TinyML at the IoT nodes however has some disadvantages.   The primary disadvantage is that bidirectional communication between the end-node and the gateway is necessary for the end-node to receive information about the characteristics of the various candidate channels, including the RSSI and SNR.  The issue with bidirectional communication is that while the gateway is engaged in transmission, it is unable to receive simultaneously, which may result in packet losses. 
The messages sent from the gateway may be minimal in their payload, which has the benefit of reducing the amount of air time required and therefor the risk of collision and of the gateway being unable to receive messages during that time. The primary objective here is to convey channel characteristics, rather than content, to the end nodes.

\begin{figure}
    \centering
    \includegraphics[width=\linewidth]{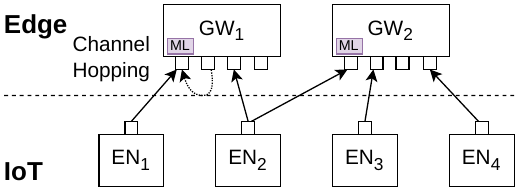}
    \caption{PLacement of TinyML on LoRa Gateway nodes (GW)}
    \label{fig:TML2}
\end{figure}

To overcome the above mentioned disadvantages, TinyML model can also be placed in the gateway, see Figure \ref{fig:TML2}. This solution obviously requires only a one-way LoRa communication channel between the IoT and edge. This is achieved through the use of RSSI and SNR values from the previous transmissions, which determine the decision to hop the channel. The most significant drawback of this approach is that when scaling the system, each end node must be flashed with a channel set, without the option of changing the used channel later. This necessitates a manual scanning of the area in which the end-node is deployed to identify interference with other channels/devices. Consequently, this process is inherently static and does not provide for enough flexibility in real world deployments.

\section{Case Study: Urban Computing Contiuum with Plant Recommender Application}\label{sec:ml}

This section presents a case study of an urban computing continuum, which incorporates the proposed LoRa channel hopping and edge computing with a plant recommendation system for urban gardening. The plant recommender chooses the most a suitable plant for an urban microfarmer based on the unique composition of each gardens soil. In a LoRa-based urban computing continuum, we incorporate the collection of sensor data in an urban area and the processing of the gathered data to develop an application that benefits the people of that area. The following subsections provide a description of the application workflow engineering and various aspects of the application in the context of IoT-edge-cloud continuum.

\subsection{Application workflow engineering} 

\begin{figure*}
    \centering
    \includegraphics[width=\linewidth]
    {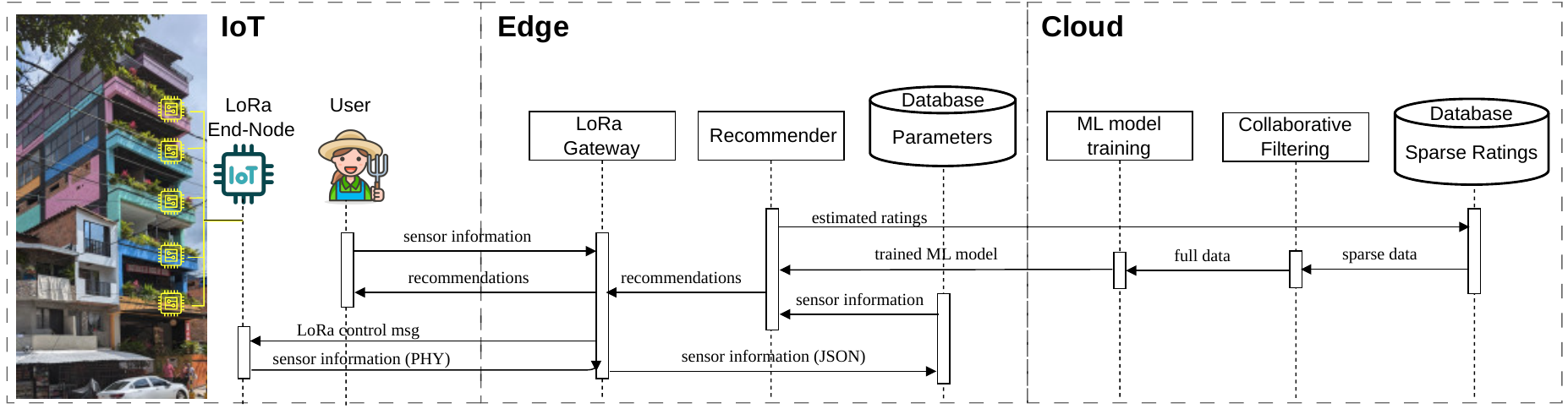}
    \caption{Engineering the application workflow, inspired by the photo from our visit in Medell\'{i}n, Colombia. }
    \label{fig:workflow}
\end{figure*}
Figure \ref{fig:workflow} illustrates the proposed application workflow with LoRa, with the assumed distance between the garden and the edge in the range of non-LPWAN protocols. The process is initiated by the urban gardener via a front-end application, which in turn initiates the transmission of a LoRa control message to the LoRa IoT end-device. Here, the soil sensor data is requested, containing the data on essential macronutrients for plant growth, \textit{Nitrogen (N)}, \textit{Phosphorus (P)}, and \textit{Potassium (K)}, all gathered along with the soil pH value and the air temperature close to the ground, via a sensor. The LoRa end-device can be a microcontroller, which is connected to the sensor and is relaying the data obtained to the edge for subsequent processing. Additionally, the urban gardener may provide data via manual ratings (e.g., scores 1 to 5) to the edge for some or all plants that have been cultivated in the previous seasons.  The LoRa gateway transforms the received information into the JSON format and subsequently relays it to the database located at the edge. The edge-based recommender then retrieves the information and forwards it to the cloud. In the cloud context, various application related data processing, such as ML training for application recommender can be performed. Here we show that a model for plants recommendation can be created either on based on complete data sets, or based on collaborative filtering for sparse data sets.  Once application model training is completed, the model is sent to the edge, where inference takes place. Based on the model, the recommender sends the plant recommendations to the user. This model is not to be confused with LoRa based channel hopping model previously discussed, as we here focus only on application workflow engineering. 

\subsection{IoT context: Data Collection}

The IoT context involves the aggregation of data. The data collection can be done via sensors, or data can also be entered manually by the user.  We assume a heterogeneous set of IoT devices on low-power, resource-constrained hardware (e.g., microcontrollers) interfacing with different sensors. This can be implemented through the use of an IP68 ABS NPK sensor in conjunction with a DS18B20 temperature sensor and a Taidda soil ph sensor, for example. It is assumed that IoT devices communicate with gateway based on an optimized LoRa physical layer communication, whereby the channel selection is performed with TinyML, as previously described (not shown here due to the focus on application). The data transmitted via LoRa comprises solely the N,P,K, pH and temperature values, in addition to an identifier for the garden, resulting in a message length of approximately 20-30 bytes; the message length needs stay as short as possible, as this reduces the overall airtime of LoRa. The sensor data is therefore transmitted without any additional information, such as spacing or explanatory text and comprises only the raw sensor values and the ID, to be processed by the edge.
\par Another option for inputting data into the system is through the manual rating of  plants. This allows for   the system to work also without deploying IoT devices. Consequently, the user is required to enter the plant ratings to the the user interface. This approach, while low-cost as it does not involve sensor hardware, introduces sparsity into the data. The sparsity of data is a known problem which in our urban continuum can be dealt with in the cloud.

\subsection{Edge context: inference and data processing}
LoRa gateway is implemented in the edge. The gateways are required to utilize the same LoRa parameters, including Coding Rate, Spreading Factor, bandwidth and base frequency employed by the IoT device. Otherwise, they will not successfully receive the packets. The purpose of the gateway is to receive the messages and add the appropriate value to the raw application data, then forward the message to the edge with which it is connected. Consequently it converts the LoRa data from bytes to integers and addresses it to the correct labels, converting the message to JSON and sending it to the application database at the edge. Furthermore, for the application, edge manages end device IDs and users, in addition to the inference of the trained application machine learning model and the provision of feedback to the end user regarding the results of the recommendation. These application related functions are preformed in addition to the LoRa system functions exectued at the edge, such as the previously described TinyML training for LoRa channel hopping optimizations.

\subsection{Cloud context: Collaborative Filtering}
In the cloud, model training is performed on either the full application data set or on sparse data, such as based on manual ranking. Prior to training the ML models on sparse data, the sparse matrices undergo pre-processing, in which they are converted into complete matrices, thereby ensuring the suitability of the data set for the training process. The missing ratings are calculated from the sparse matrix using cosine similarity, thereby generating a complete ratings matrix for all listed plant and soil combinations. In the context of recommendation systems, the cosine similarity metric is calculated using the ratings of two items, in our use case these are the ratings for the plants, designated as "x" and "y", which depict two rows in the sparse matrix. The ratings on each soil represent the input data, an the cosine similarity is given by
\begin{equation}
\cos(x,y) = \frac{x \cdot y}{\|x\| \|y\|}
\end{equation}

The sparse matrix $S \in \mathbb{N}^{m \times n}$ is composed of natural numbers, e.g.,  $\in \Iintv{1,5}$, representing the ratings, in addition to missing ratings situated between these values. We define the sparsity as the percentage of missing ratings. The parameters $m$ and $n$ describe the amount of soil samples and the number of plants that can be rated, respectively. Cosine similarity  is used to compare all $m$ rows with each other, to construct the full matrix $F \in \mathbb{N}^{m \times n}$. In cases where data is absent for a given entry, the sparse matrix is populated by identifying similar soils based on the cosine similarity and calculating a weighted average of the ratings assigned to these soils based on their degree of similarity. Subsequently, the newly calculated ratings are incorporated into the sparse matrix, thereby generating a comprehensive data set of ratings.
\par Finally, the data resulting from the calculation of the cosine similarity is employed to train the actual recommendation system in the cloud. In our applicatoin,  we choose the supervised learning models  K-Nearest Neighbors (KNN), Neural Networks (NN), Linear Regression, Decision Tree, Random Forest, as well as the ensembled learning models Gradient Boost (GB) and Extreme Gradient Boost (XGB), due to their capacity to perform regression tasks and their widespread adoption, in addition to the availability of the related libraries. We deployed these models for the plant recommendation application which we next analyze in terms of accuracy, error rate, training time and inference time. 



\section{Experimental and Simulation Results}\label{sec:results} 
 This section shows experimental and simulation results, including a comparative analysis of the efficacy of low-power communication solution (LoRa) in the context of the proposed TinyML assisted channel hopping solutions to transmit sensor data from the IoT device to the edge with minimized collisions. We also show application related performance results, based on simulations, in the context of collaborative filtering for addressing incomplete data. The parameters utilized in experiments and simulations are outlined in Table \ref{tab:table2}.



\begin{table}[h!]
  \begin{center}
    \caption{Experimental and Simulation Parameters} 
    \label{tab:table2}
    \begin{tabular}{|c|c|c|}
     \specialrule{.15em}{0em}{0em} 
      \textbf{Experiment/ } &    \textbf{Parameters} & \textbf{Values}  \\
      \textbf{ Simulation} &     &   \\

      \specialrule{.15em}{0em}{0em} 
      LoRa & Distance & 1 m , 15 m , 30 m \\
      (Experiment) & Frequency  &868.0, 869.0, 870.0MHz\\
        & Bandwidth  & 125 kHz \\
        & Coding Rate  & 4/5 \\
        & Spreading Factor  & 7 \\
        & MCU End Node & Heltec WiFi LoRa 32 (V3) \\
        & LoRa radio chip & SX1262 \\
        & MCU gateway & Adafruit Feather M0 \\
        & LoRa radio chip & SX1276 \\
      \hline
       TinyML  &  library  & tensorflow  \\
        (Simulation &    & $\rightarrow$ tensorflow.keras \\
        \& experiment) &    & TensorFlowLite\_ESP32.h \\
           & model & Sequential: Dense(10) \\
             &  & Dense(10,  regularizer=L1) \\
             &  & Dense(F, activ.='softmax') \\
             &   & optimizer=Adam  \\  
             & Data size & 5000 for 1 end-node \\
      \hline
       Collaborative  &  library  & scikit-learn  \\
        filtering &    & $\rightarrow$ cosine\_similarity \\
        (Simulation)   & Number of plants  & 100 \\
         & Number of soils  & 22013  \\  
         \hline
       Plant  recommender & Model library  & scikit-learn \\
        (Simulation) & Number of soils  & 22013 \\
         & Number of plants  & 100 \\
          & test size data  & 20 \% \\
      \specialrule{.15em}{0em}{0em} 
    \end{tabular}
  \end{center}
\end{table} 
\subsection{LoRa performance }
\subsubsection{Experimental Setup}
\begin{figure}
    \centering
    \includegraphics[width=\linewidth]{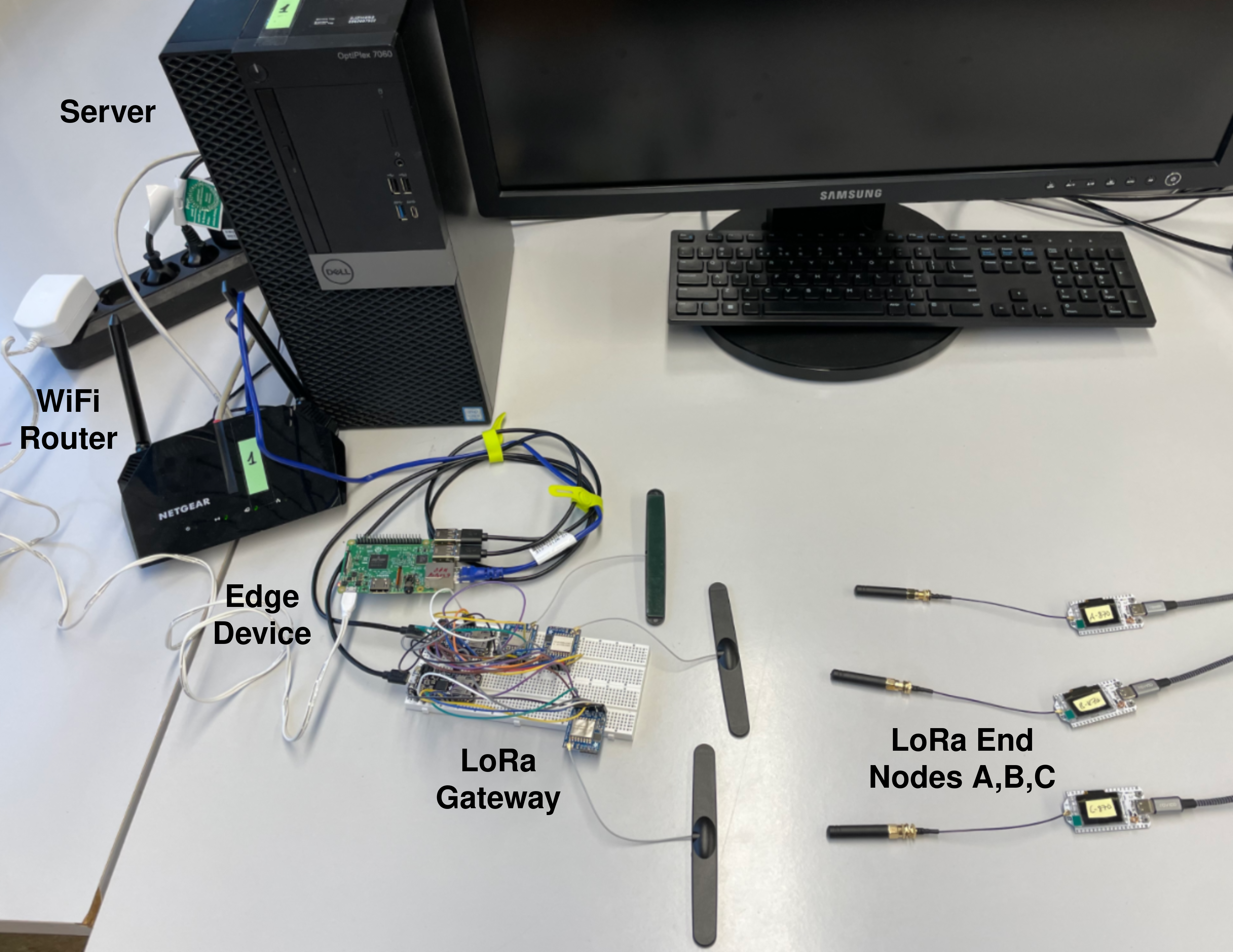}
    \caption{All components of communication workflow.}
    \label{fig:setup}
\end{figure}

\begin{figure}
    \centering
    \includegraphics[width=\linewidth]{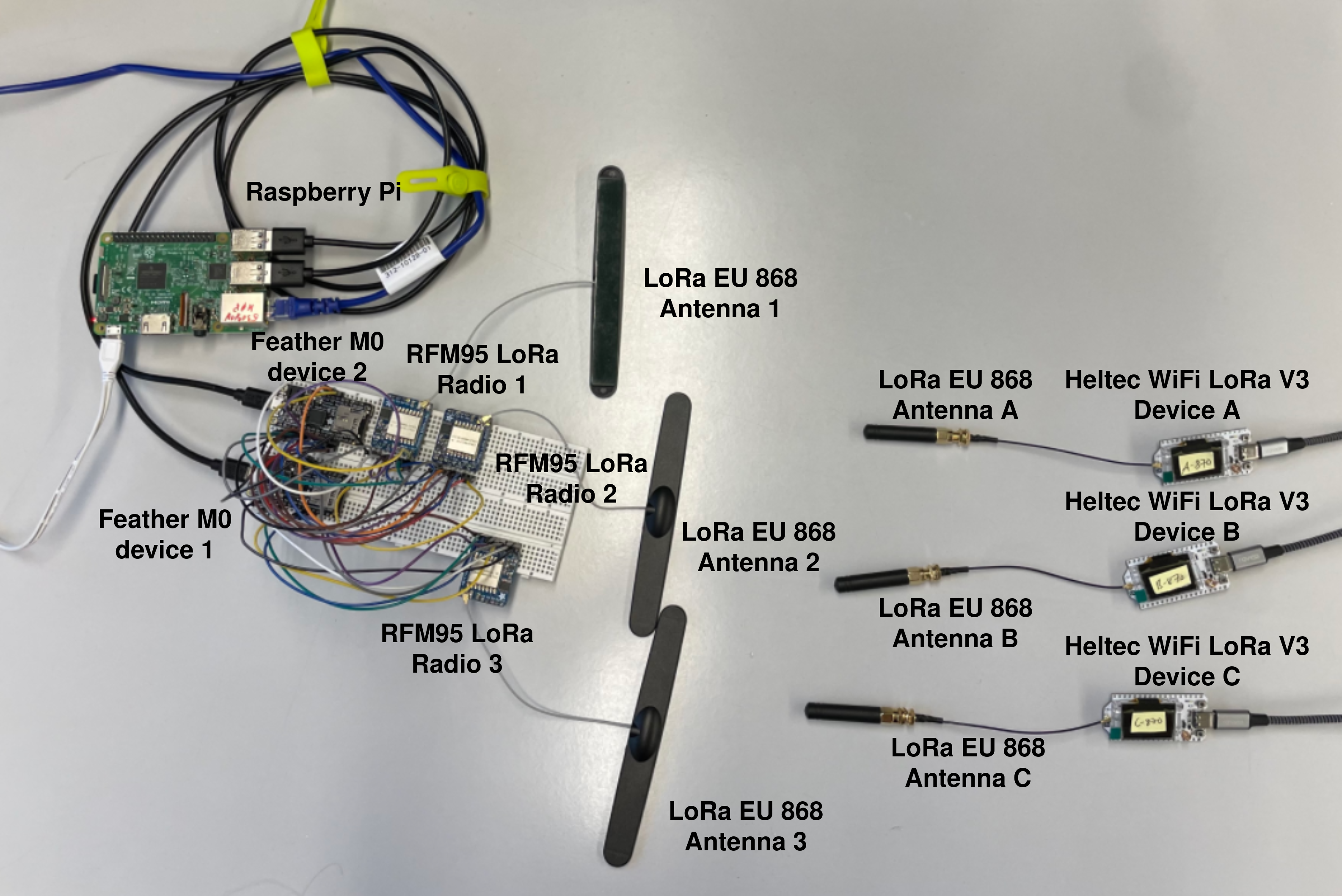}
    \caption{LoRa gateway/edge and IoT end-nodes}
    \label{fig:gwen}
\end{figure}

\begin{figure}
    \centering
    \includegraphics[width=0.95\linewidth]{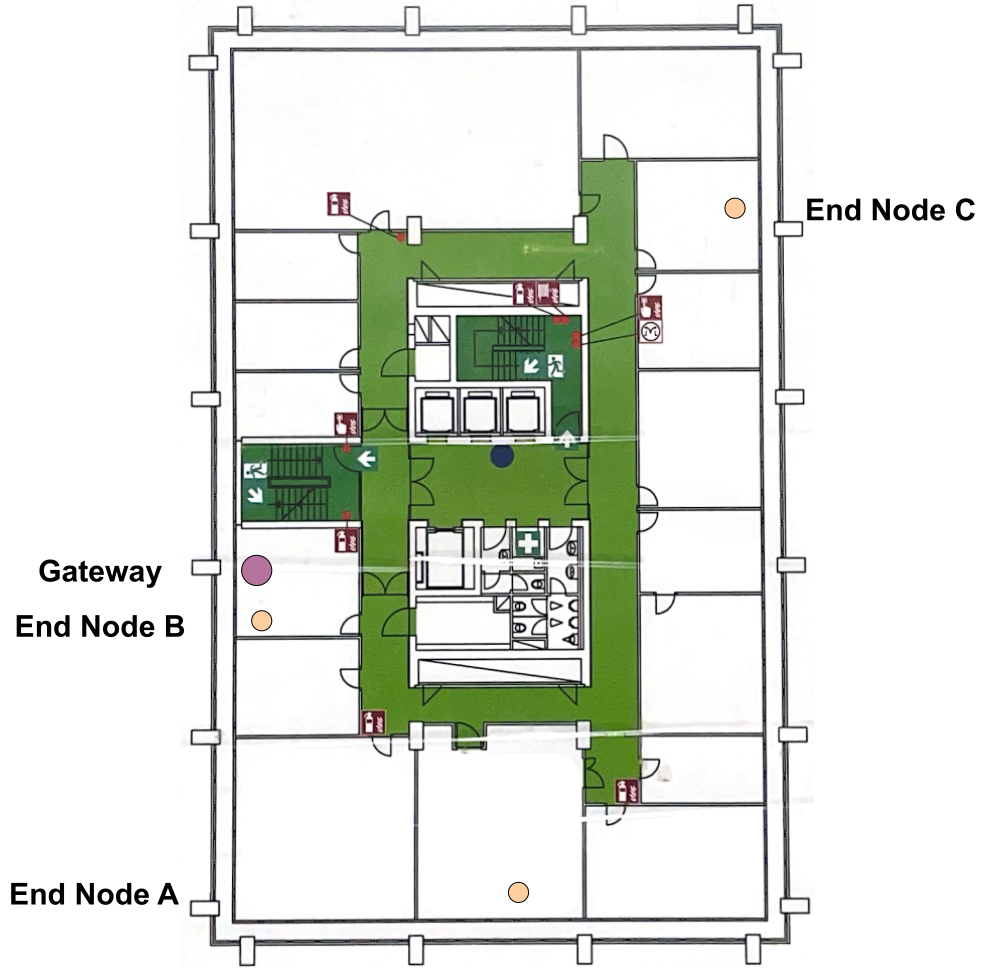}
    \caption{Position of LoRa IoT End Nodes and Edge gateway in the experimental setup.}
    \label{fig:floorplan}
\end{figure}

We implement an experimental setup to be able to measure the LoRa performance to evaluate the effectiveness of the proposed channel hopping algorithm for edge computing. The setup is used as baseline implementation for application performance results, shown later. 
The experiments were conducted by placing three LoRa IoT end-nodes (denoted as Devices A, B and C) in a  laboratory setup.
The hardware of the end devices comprises a Heltec WiFi Lora 32 V3 which features an integrated LoRa SX1262 chip and an antenna optimized for the EU686 band. 
THe gateways are two Adafruit Feather M0 microcontrollers, which are connected to a total of three RFM95W LoRa radios, running on SX1276 LoRa chips. Using these three radios, the gateway is able to receive and log LoRa packets on three different frequencies simultaneously. The gateways are coupled with a Raspberry Pi which acts as the edge computing device. The experimental setup presenting all components of the experiment are shown in Figure \ref{fig:setup} with a detailed overview of the hardware used for LoRa gateway and LoRa end-node in Figure \ref{fig:gwen}. In all experiments, the spreading factor was set to 7, the bandwidth to 125 kHz, the coding rate to 4/5 and the distance between LoRa gateway and end nodes (A,B,C) was 15, 1 and 30 meters respectively. The precise positioning of the devices is illustrated in Figure \ref{fig:floorplan}.
To conduct the experiments,  three LoRa end-nodes where flashed with an identical code, except for a different identifier at the beginning of each message, in order to maintain separation between the sending nodes at the gateway. The uniform code serves to ensure that the periodic sending rhythm between them is not disrupted, as LoRa is typically deployed in periodic cases, which we are attempting to mimic here. All three nodes were activated within a two-second time interval, in comparison to the other nodes. Each node transmits 50 LoRa messages with a five-second interval between each transmission, after which the message size is increased. This is archived by utilizing a starting payload length of 30, which is incrementally increased to a maximum length of 250 bytes in 5 steps. Upon transmitting the last 250 byte long packet, the individual IoT end-nodes undergo a change in the base frequency. This process is repeated twice, resulting in the coverage of the three distinct frequencies. 

\par In all experiments, the following metrics were measured at the gateway: RSSI,SNR and packet delivery ratio (PDR). The results are shown in Tables \ref{tab:868}, \ref{tab:869}, \ref{tab:870}. The results demonstrate that since the device B was located in the closest proximity to the gateway (1m), it achieves the best RSSI values measured across all conducted experiments. Similarly, it can be observed that device C was positioned at the farthest distance, while device A was situated in the intermediate range. In the context of this experiment, carried out in a given location at a designated point in time, 869 MHz frequency was measured to be the most reliable one, i.e., with the lowest packet loss across all three devices. Frequencies 868 MHz and 870 MHz demonstrated inconsistent performance, with only about 10 \% of packets reaching the gateway in certain areas of the building. This is despite the relatively limited communication range, which should not have posed a significant challenge for LoRa.
Table \ref{tab:868} and table \ref{tab:870} show a surprising phenomenon: The device C with the lowest RSSI and SNR, which is the farthest from the other devices, exhibits the best performance. The observed effect can be attributed to interference on the frequency band used. This may originate from the concurrent transmission of LoRa packets by the neighboring nodes, or from external devices occupying the same frequencies. It could also stem from the simultaneous arrival of two LoRa packets at the gateway, which may cause the weaker of the two signals to fail to be demodulated. It is challenging to pinpoint the precise source, location and time of the packets lost. The number of packets transmitted and received along with the specific packets involved are the only values that can be measured.
This demonstrates in fact the need to optimizing channel selection in order to guarantee reliable, as we show next. 

%
%
\begin{table}[h!]

  \begin{center}
    \caption{Measurements for 3 LoRa end-nodes at 868 MHz}
    \label{tab:868}
\begin{tabular}{lccccr}
\hline
 source   &   size of message&     RSSI &     SNR &   count &  PDR \\
     &  [KB] &     &     &    &    \\
\hline
 A        &                30 &  -90.1 & 9.1  &       9 &                  0.18 \\
 A        &                74 &  -91.2 & 9.5     &       6 &                  0.12 \\
 A        &               118 &  -86.5    & 9.5     &       4 &                  0.08 \\
 A        &               162 &  -88.5    & 9.3    &       4 &                  0.08 \\
 A        &               206 &  -89.5    & 9       &       2 &                  0.04 \\
 A        &               250 &  -86.6 & 9       &       7 &                  0.14   \\
 \hline
  B        &                30 &  -44.6  & 9.1   &       8 &                  0.16 \\
 B        &                74 &  -43.2   & 9.4    &       4 &                  0.08 \\
 B        &               118 &  -42.6 & 9.7 &       3 &                  0.06 \\
 B        &               162 &  -43.6 & 9.7 &       3 &                  0.06 \\
 B        &               206 &  -42.5    & 9.5     &       2 &                  0.04 \\
 B        &               250 &  -76.3 & 3.3    &      12 &                  0.24    \\
\hline
 C        &                30 & -112.1  & 4.3 &      49 &                  0.98 \\
 C        &                74 & -110.6    & 5.4    &      50 &                  1    \\
 C        &               118 & -110.4   & 5.4    &      50 &                  1    \\
 C        &               162 & -110.8   & 5.1    &      50 &                  1    \\
 C        &               206 & -111.1   & 4.9    &      50 &                  1    \\
 C        &               250 & -112.3   & 4       &      50 &                  1    \\
\hline
\end{tabular}
  \end{center}
\end{table} 

\begin{table}[h!]
  \begin{center}
    \caption{Measurements for 3 LoRa End-nodes at 869 MHz}
    \label{tab:869}

\begin{tabular}{lccccr}
\hline
 source   &   size of message&     RSSI &     SNR &   count &  PDR \\
     &  [KB] &     &     &    &    \\

\hline
 A        &                30 &  -71.5   & 9.3    &      50 &                  1    \\
 A        &                74 &  -71.7  & 9.3    &      50 &                  1    \\
 A        &               118 &  -71.6   & 9.1     &      50 &                  1    \\
 A        &               162 &  -71.4   & 9.1     &      50 &                  1    \\
 A        &               206 &  -70.3  & 9.4     &      50 &                  1    \\
 A        &               250 &  -70.1  & 9.2     &      50 &                  1    \\
 \hline

 B        &                30 &  -37.9   & 9.5    &      50 &                  1    \\
 B        &                74 &  -35.8   & 9.5    &      50 &                  1    \\
 B        &               118 &  -34.7  & 9.5    &      50 &                  1    \\
 B        &               162 &  -36.1  & 9.1    &      50 &                  1    \\
 B        &               206 &  -34.1  & 9.3     &      50 &                  1    \\
 B        &               250 &  -35.1  & 9.1    &      50 &                  1    \\
 \hline
 C        &                30 & -108.1  & 7.5    &      50 &                  1    \\
 C        &                74 & -108.9 & 6.8  &      49 &                  0.98 \\
 C        &               118 & -115.5  & 3.2    &      50 &                  1    \\
 C        &               162 & -117.9 & 2.0 &      46 &                  0.92 \\
 C        &               206 & -114.0     & 4.5    &      50 &                  1    \\
 C        &               250 & -118.1 & 1.4 &      44 &                  0.88 \\
\hline
\end{tabular}
  \end{center}
\end{table}

\begin{table}[h!]
  \begin{center}
    \caption{Measurements for 3 LoRa end-nodes at 870 MHz}
    \label{tab:870}

\begin{tabular}{lccccr}
\hline
 source   &   size of message&     RSSI &     SNR &   count &  PDR \\
     &  [KB] &     &     &    &    \\
\hline
 A        &                30 &  -70.0      & 9.2 &       9 &                  0.18 \\
 A        &                74 &  -75.5    & 9       &       6 &                  0.12 \\
 A        &               118 &  -72.8   & 9       &       4 &                  0.08 \\
 A        &               162 &  -72.8   & 9.3    &       4 &                  0.08 \\
 A        &               206 &  -80.0      & 9.5     &       2 &                  0.04 \\
 A        &               250 &  -80.3   & 9       &       4 &                  0.08 \\
 \hline
 B        &                30 &  -32.1 & 9       &       7 &                  0.14 \\
 B        &                74 &  -29.8    & 9.8     &       5 &                  0.1  \\
 B        &               118 &  -29.0      & 9.3 &       3 &                  0.06 \\
 B        &               162 &  -30.4 & 9       &       7 &                  0.14 \\
 B        &               206 &  -34.3 & 9.3 &       3 &                  0.06 \\
 B        &               250 &  -34.0      & 9       &       3 &                  0.06 \\
 \hline
 C        &                30 & -101.1   & 4.4    &      50 &                  1    \\
 C        &                74 & -100.5  & 4.2 &      49 &                  0.98 \\
 C        &               118 & -102.1   & 2.3    &      50 &                  1    \\
 C        &               162 & -104.8  & 1       &      46 &                  0.92 \\
 C        &               206 & -106.1   & 5.9 &      49 &                  0.98 \\
 C        &               250 & -113.4   & 4.6     &      50 &                  1    \\
\hline
\end{tabular}
  \end{center}
\end{table} 

\subsubsection{TinyML Performance}
\begin{figure}
    \centering
    \includegraphics[width=\linewidth]{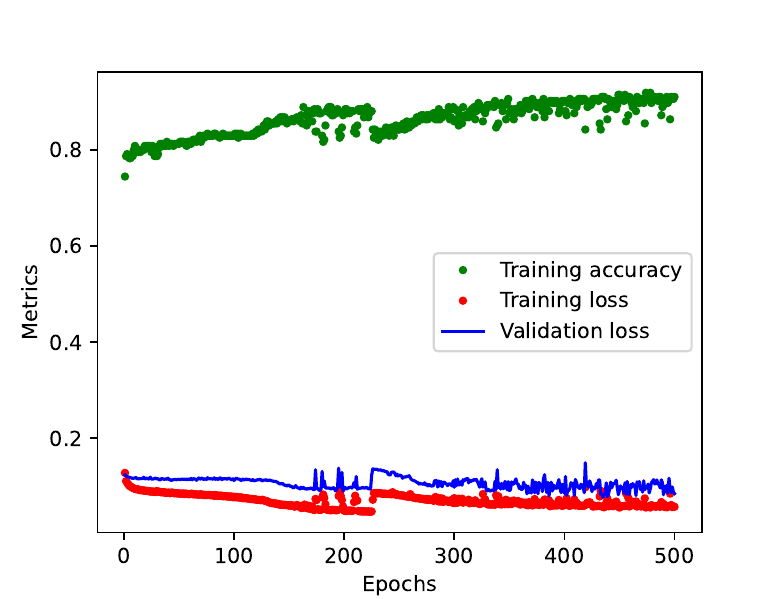}
    \caption{TinyML model Performance Trends: Accuracy, Training Loss, and Validation Loss by Epoch.}
    \label{fig:training}
\end{figure}

The parameters used in the TinyML experiment are defined in  Table \ref{tab:table2}. The model has two Dense layers with  L1 kernel regularizer, a softmax activation function at the output layer with size equal to the number of channels, and Adam optimizer. We divided the data into training, validation and testing, with 60\%, 20\%, 20\% distribution, respectively. 
Figure \ref{fig:training} presents the accuracy and validation results after the training process. We can observe a consistent decrease in loss alongside an increase in accuracy as training progresses. This trend indicates that the model successfully learned patterns from the input sequence data, enabling it to improve its predictions and achieve higher accuracy. 

\begin{figure*}
 \centering 
 \begin{subfigure}[t]{0.30\linewidth}
 \includegraphics[scale=0.35]{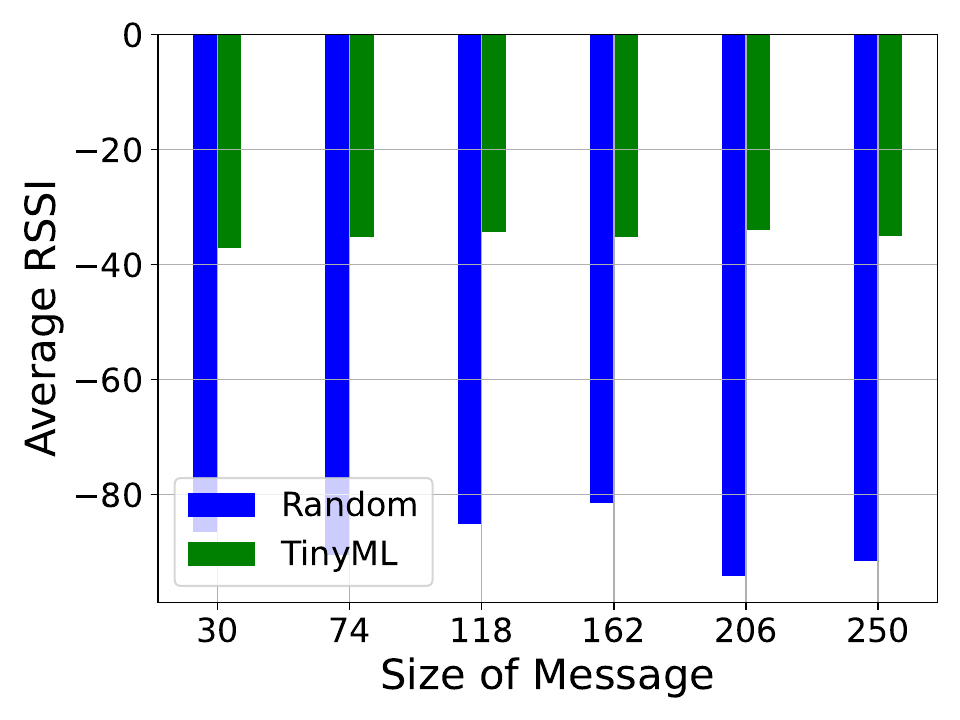}
  \caption{Average RSSI }\label{subfig:RSSI}
   \end{subfigure}
   \begin{subfigure}[t]{0.30\linewidth}
 \includegraphics[scale=0.35]{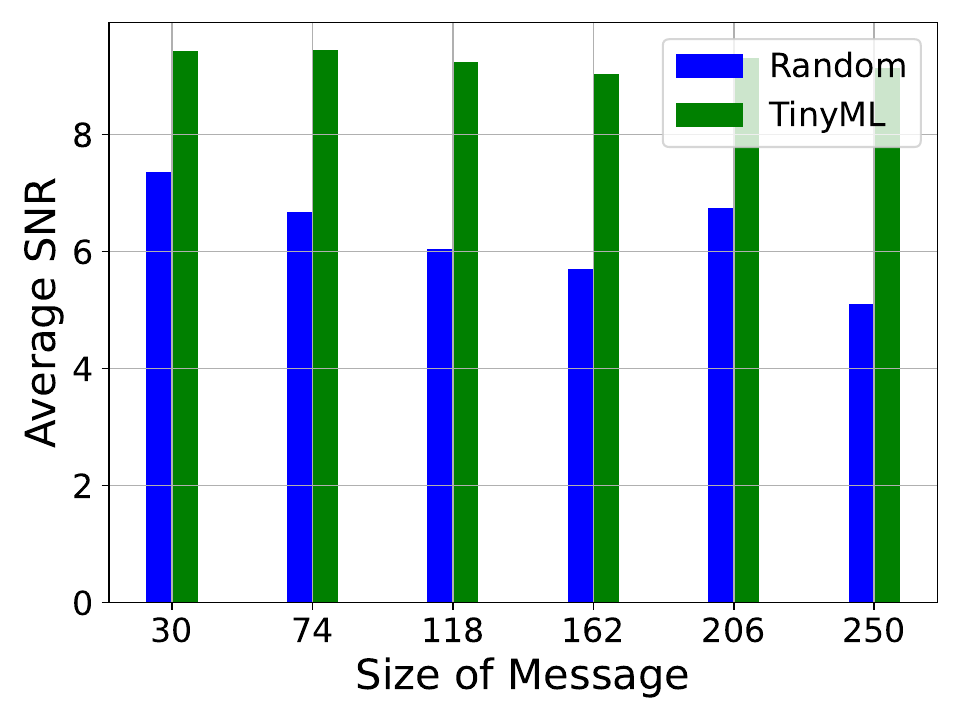}
 \caption{Average SNR }\label{subfig:SNR}
   \end{subfigure}
 \begin{subfigure}[t]{0.30\linewidth}
 \includegraphics[scale=0.35]{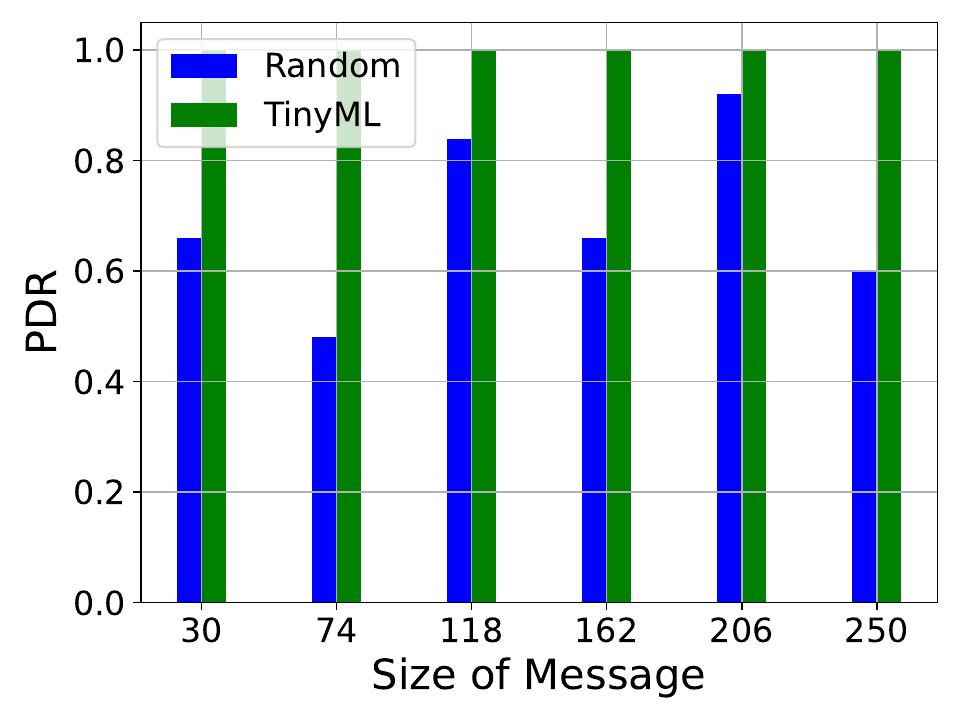}
 \caption{PDR  }\label{subfig:PDR}
   \end{subfigure}
 \caption{Performance measurements: Random channel hopping vs. TinyML}
\label{fig:TMLPerf}
\vspace{-0.2 cm}
\end{figure*}

After training, the TinyML model is used to predict the future channel to hop, using the input described previously, and then compared to a random guess approach in terms of average RSSI, average SNR, and PDR. Figure \ref{fig:TMLPerf} shows the results of this comparison. The RSSI values demonstrate a notable improvement, reaching up to 63 \% better values for a message size of 206 bytes in comparison to the randomly hopping method. Similarly, the SNR values exhibit a considerable improvement, reaching up to 44 \% better values. With regard to the PDR, the TinyML algorithm is capable of identifying channels that can successfully transmit all packets. This is a consequence of the data used for training, presented in Tables \ref{tab:868} to \ref{tab:870}, which demonstrate that for each source node and frequency, there was always a channel that can transmit all packets successfully. As a result, the TinyML model is capable of effectively selecting the appropriate channels for transmission, thereby demonstrating its efficacy.

In order to run TinyML model in the microcontroller, we first need to convert it to TFLite format (*.tflite), as shown in the Listing \ref{lst:conv}. 
\begin{lstlisting}[language=PythonCustom, caption={Python script for TFLite model conversion}, label={lst:conv}]
import tensorflow as tf
converter = tf.lite.TFLiteConverter\
              .from_keras_model(model)
tflite_model = converter.convert()
# Save the model to disk
open("hopping_model.tflite", "wb").write(tflite_model)
\end{lstlisting}

Afterwards, TFLite model is converted to a C format (*.h file) to be included in the c files used to flash the microcontroller. The TFLite to C conversion is shown in Listing \ref{lst:Cconv}.

\begin{lstlisting}[language=PythonCustom, caption={TFLite to C model conversion}, label={lst:Cconv}]
import subprocess
# Write the start of the header file
with open("model.h", "w") as f:
    f.write('const unsigned char model[] = {\n')
# Append the .tflite file converted to C 
subprocess.run("cat hopping_model.tflite | xxd -i >> model.h", shell=True, check=True)
# Finish the header file
with open("model.h", "a") as f:
    f.write('};\n')
\end{lstlisting}

\begin{figure}
    \centering
    \includegraphics[width=\linewidth]{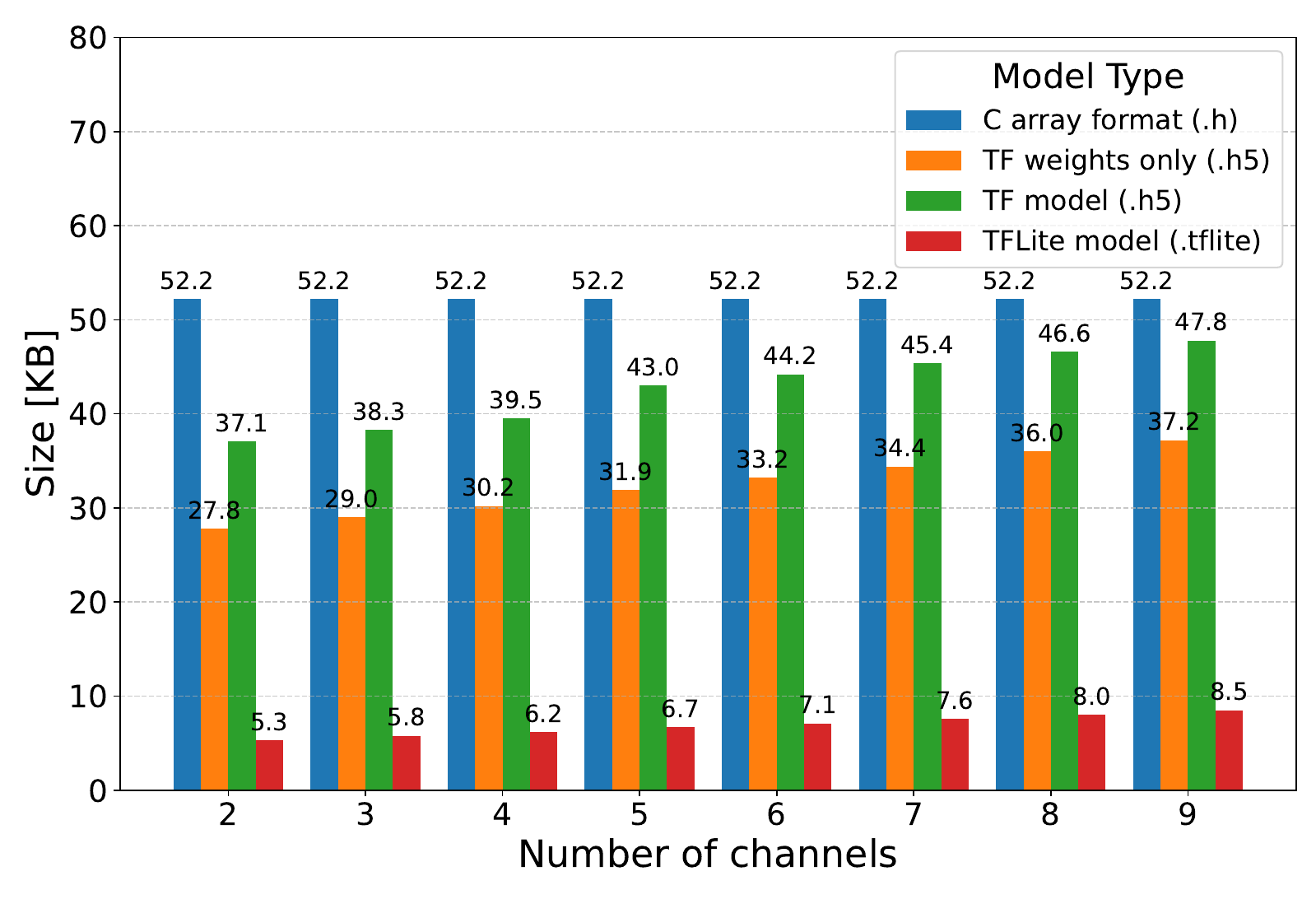}
    \caption{TinyML model size comparison across different channel hopping options.}
    \label{fig:modelsizes}
\end{figure}

Figure \ref{fig:modelsizes} compares the size in kilobytes [KB] of four formats of the proposed ML model: C array format (.h), TF weights only (.h5), TF model (.h5), TFLite model (.tflite), considering different channel hopping options, i.e., number of channels that the end-node can use. We note that the TF model is much higher than the TFLite model for all channel hopping scenarios (from hopping between two channels to hopping between 9 channels). For instance, for 2-channel hopping the TF model has a size of 37.1 KB while the TFLite model size  is 5.3 KB; 9-channel hopping the TFLite model size is 8.5 which is smaller than the TF model. By converting the TFLite model into a C array format, we obtained a file with size 52.2 KB for all hopping options, which is an acceptable size to be used in the microcontroller ESP32.

\subsection {Application Performance}
For the baseline configuration shown previously and implemented experimentally, we now show the application performance which we obtain by simulations.
\subsubsection{Collaborative filtering performance} 
\begin{figure*}[htbp]
    \centering 
    \begin{subfigure}[t]{0.31\linewidth}
        \includegraphics[scale=0.22]{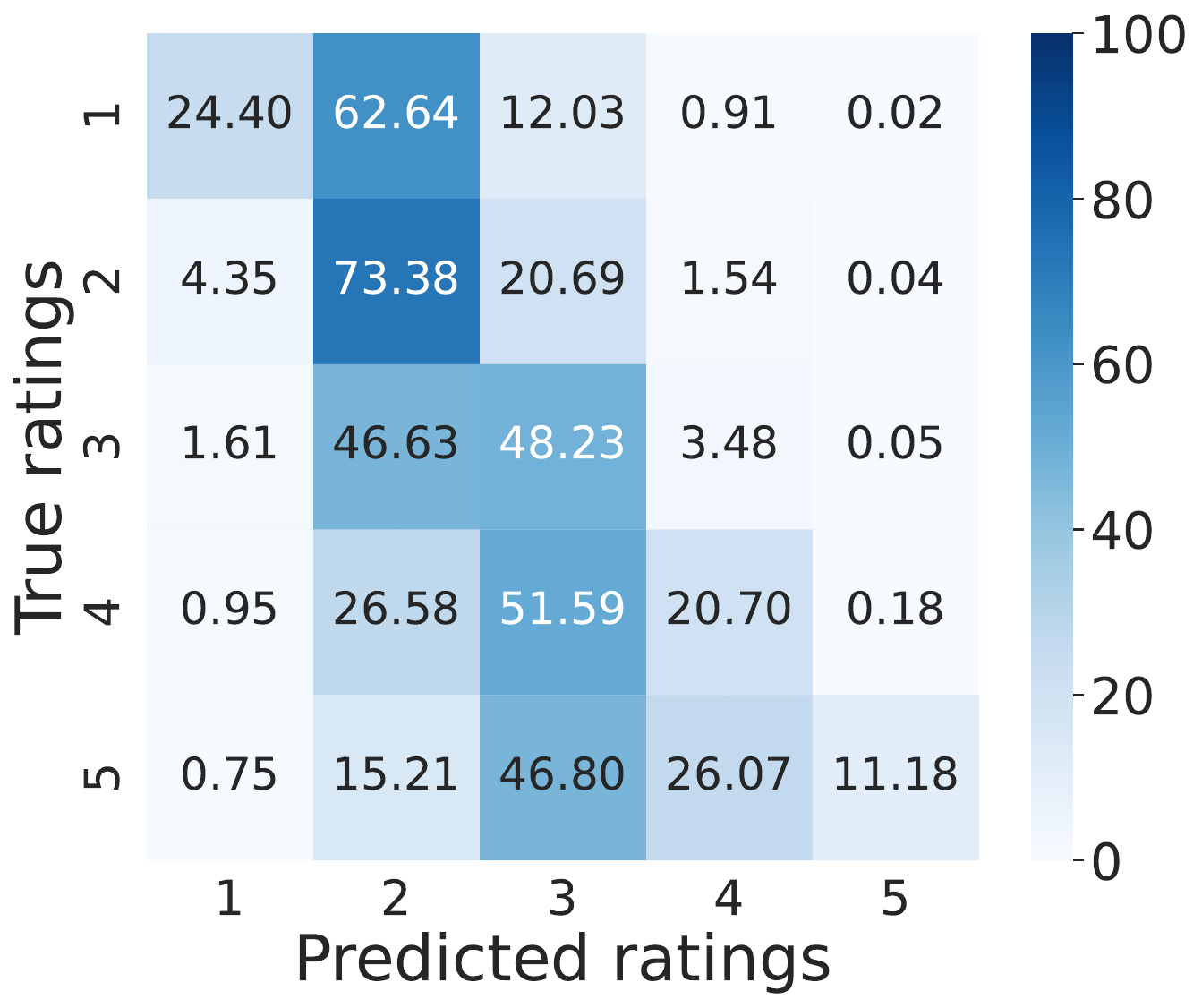}
        \caption{Sparsity=90\%}
        \label{subfig:mat1}
    \end{subfigure}
    \begin{subfigure}[t]{0.31\linewidth}
        \includegraphics[scale=0.22]{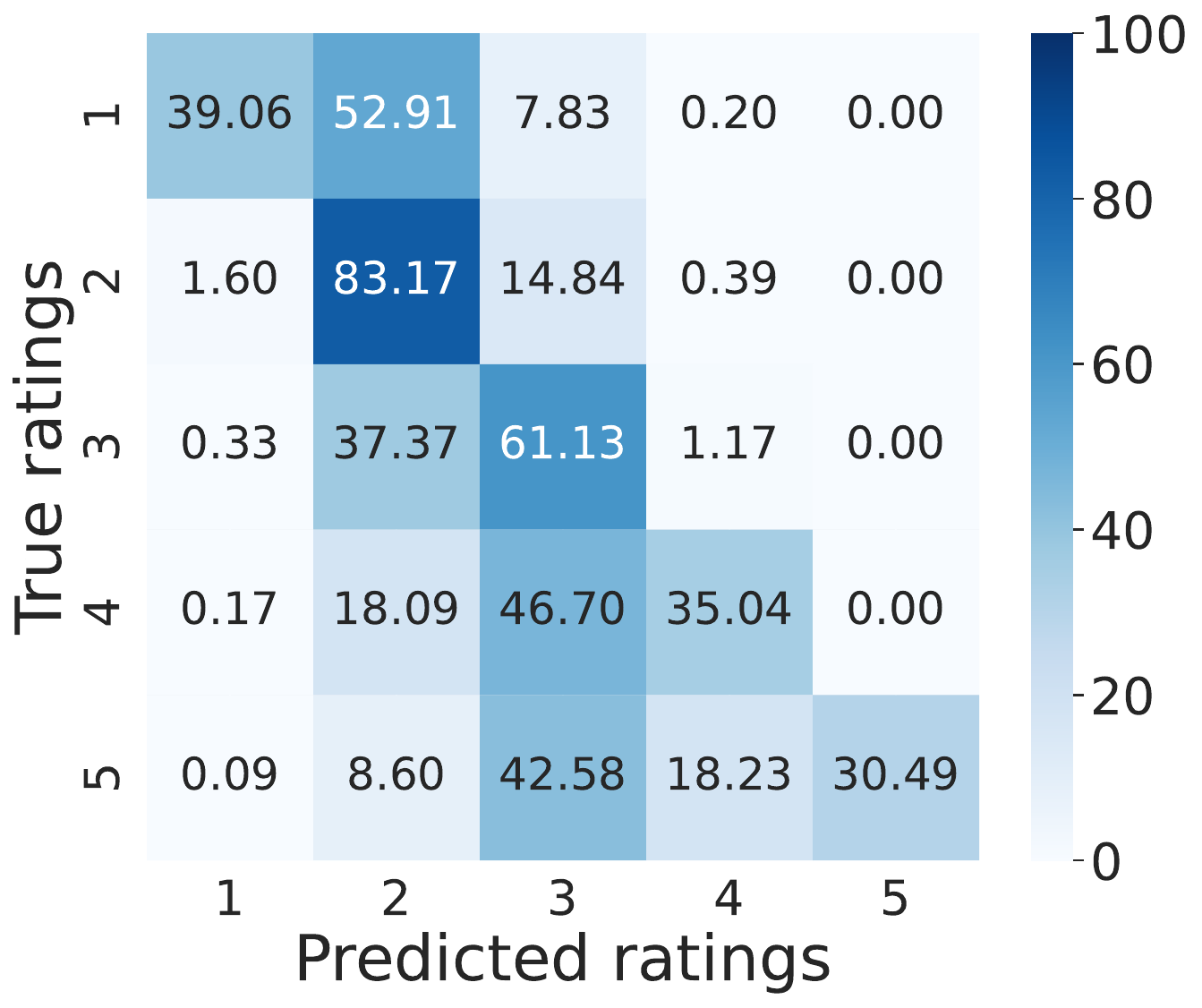}
        \caption{Sparsity=70\%}
        \label{subfig:mat2}
    \end{subfigure}
    \begin{subfigure}[t]{0.31\linewidth}
        \includegraphics[scale=0.22]{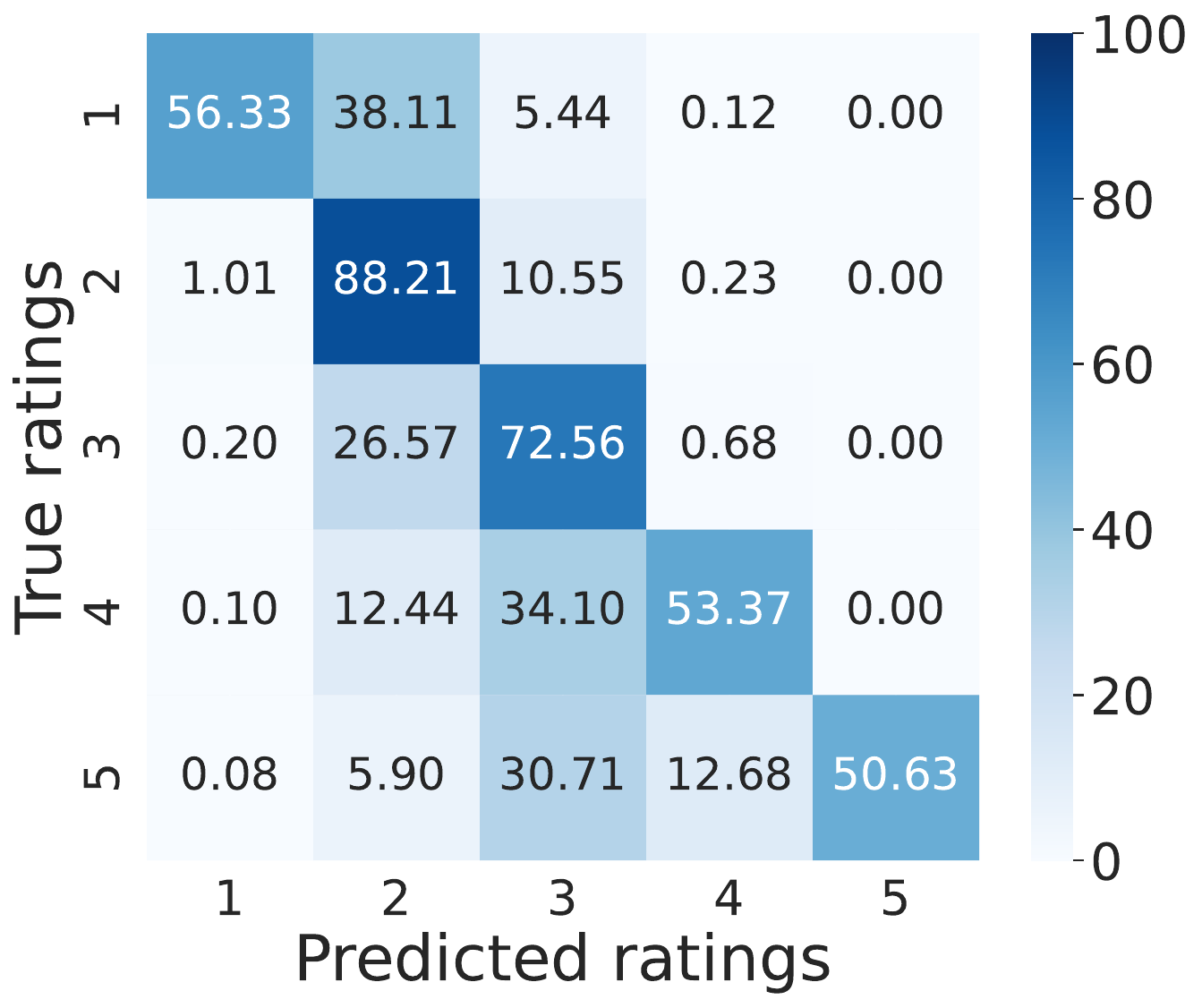}
        \caption{Sparsity=50\%}
        \label{subfig:mat3}
    \end{subfigure}

    \vspace{0.1cm}

    \begin{subfigure}[t]{0.31\linewidth}
        \includegraphics[scale=0.22]{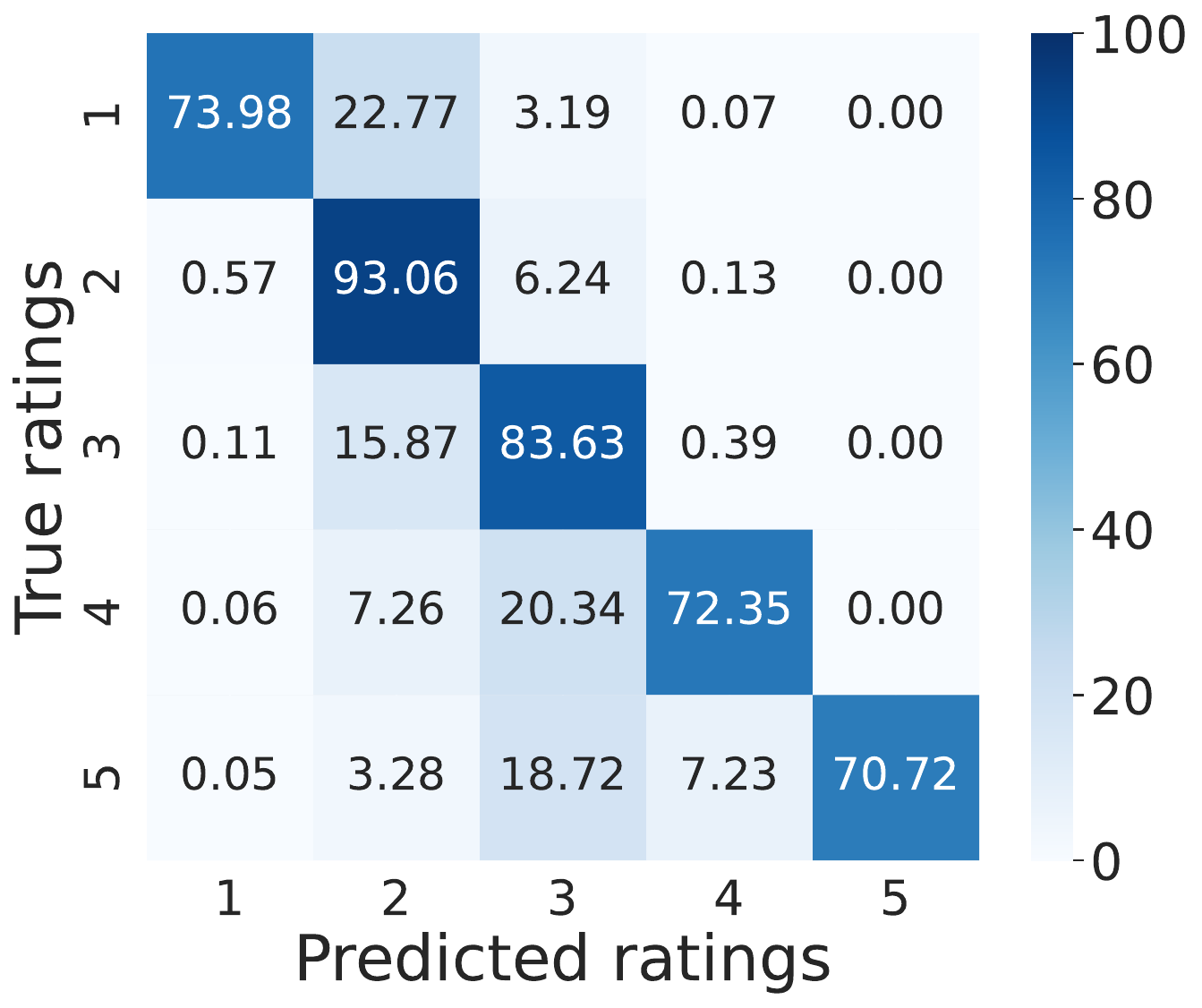}
        \caption{Sparsity=30\%}
        \label{subfig:mat4}
    \end{subfigure}
    \begin{subfigure}[t]{0.31\linewidth}
        \includegraphics[scale=0.22]{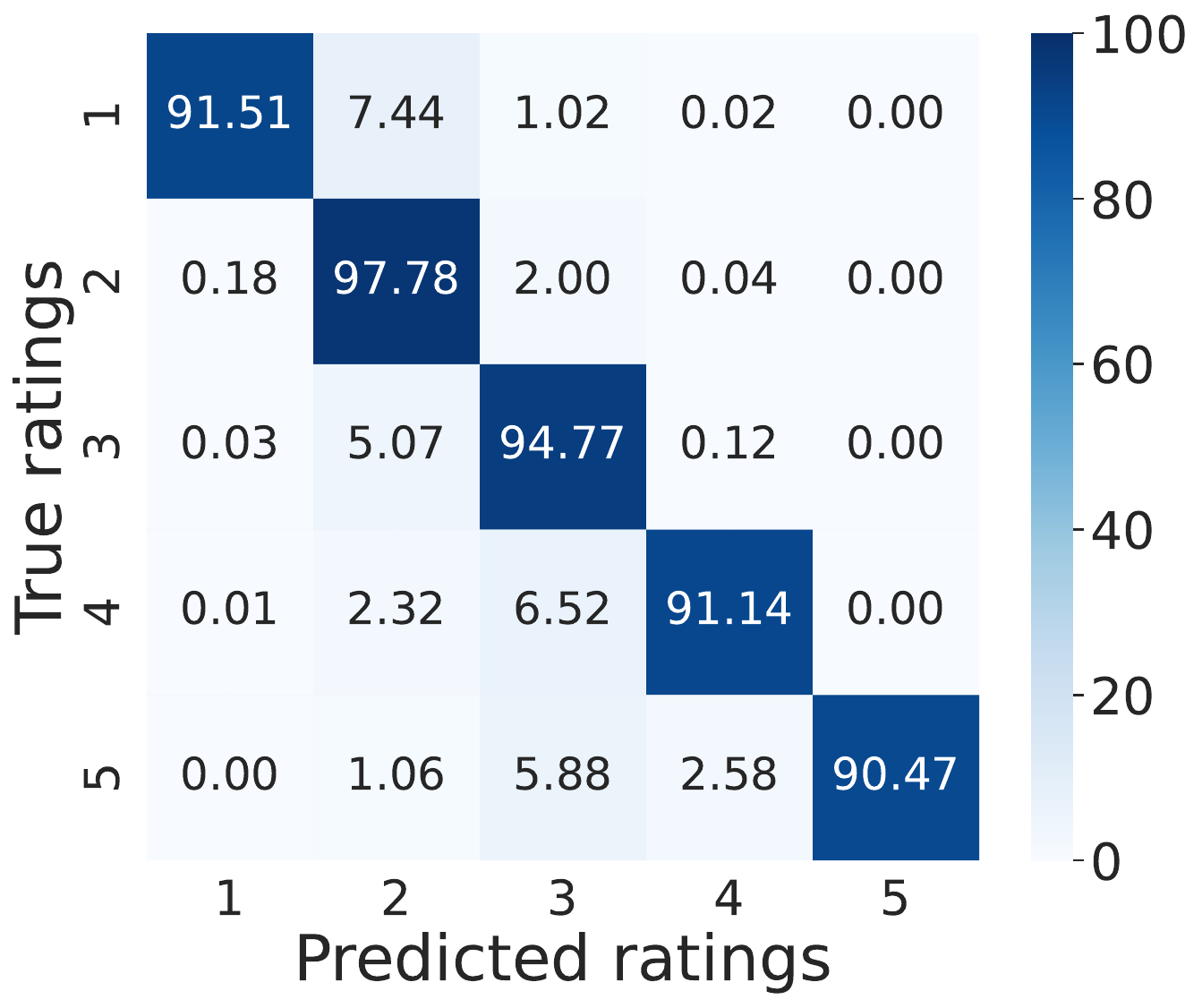}
        \caption{Sparsity=10\%}
        \label{subfig:mat5}
    \end{subfigure}
    \begin{subfigure}[t]{0.31\linewidth}
        \includegraphics[scale=0.22]{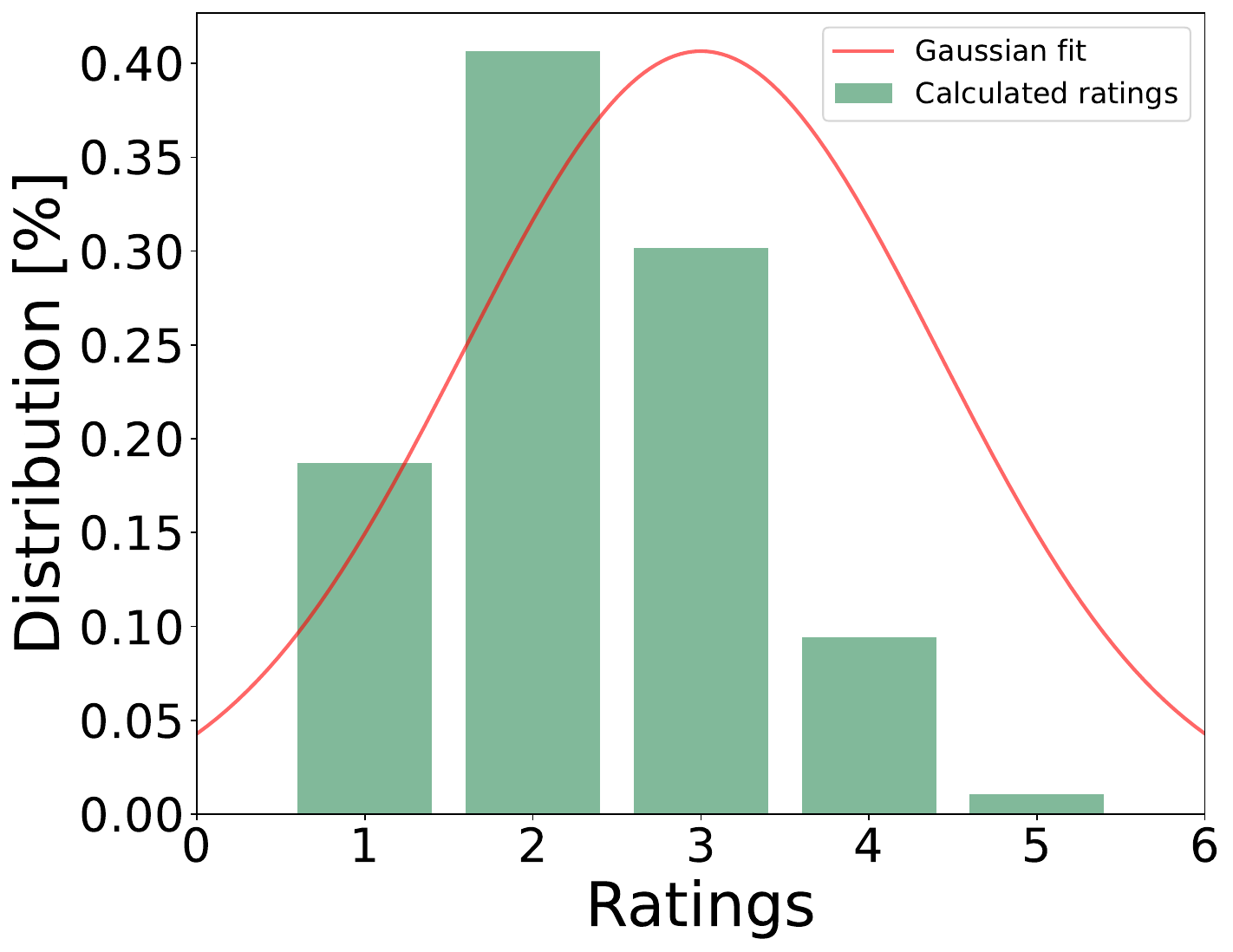}
        \caption{Distribution}
        \label{subfig:mat6}
    \end{subfigure}

    \caption{Confusion matrices and distribution with different applicatoin data sparsity.}
    \label{fig:combined_confmat}
    \vspace{-0.2cm} 
\end{figure*}

We define application data sparsity as the percentage of missing ratings. To analyze its impact on the recommender system, we implement the following test.  First, a application dataset with all ratings is calculated based on optimal conditions of the soils, here an application dataset comprising of 22000 soil samples on 100 plants. In the following step, i.e., data filtering, 10\%, 30\%, 50\%, 70\% and 90\% of the ratings are removed. Subsequently, we apply cosine similarity on this filtered dataset to identify soil samples that exhibit similar characteristics based on the remaining ratings; we calculate the cosine similarity of every pair of different soil data. Based on the most similar soils, we calculate a weighed average of the non-zero ratings to fill in for the missing ratings of a soil. To compare this data with the original dataset, the calculated ratings are rounded to the nearest integer. The results are shown in Figures \ref{subfig:mat1}-\ref{subfig:mat5}, which analyze the efficacy of the cosine similarity metric on the given data set by constructing a confusion matrix. As sparsity levels decrease, the comparison becomes increasingly unfavorable. This is due to the reduction in the quantity of data available to calculating similarity and corresponding ratings. The degree of confusion is notably elevated at sparsity levels 90 \% to 50 \%, as shown in Figure \ref{subfig:mat6}, which shows the distribution of ratings across the matrix. This result illustrates that the majority of the ratings fall within the range of 1-3, suggesting that they have a strong impact on the cosine similarity performance. At sparsity level 10, i.e., corresponding to 10 \% of missing data, the ratings could be predicted with a minimum of 90 \% accuracy across all rating categories. 

\subsubsection{Plant recommendations} 

\begin{figure}
 \centering 
   \includegraphics[scale=0.4]{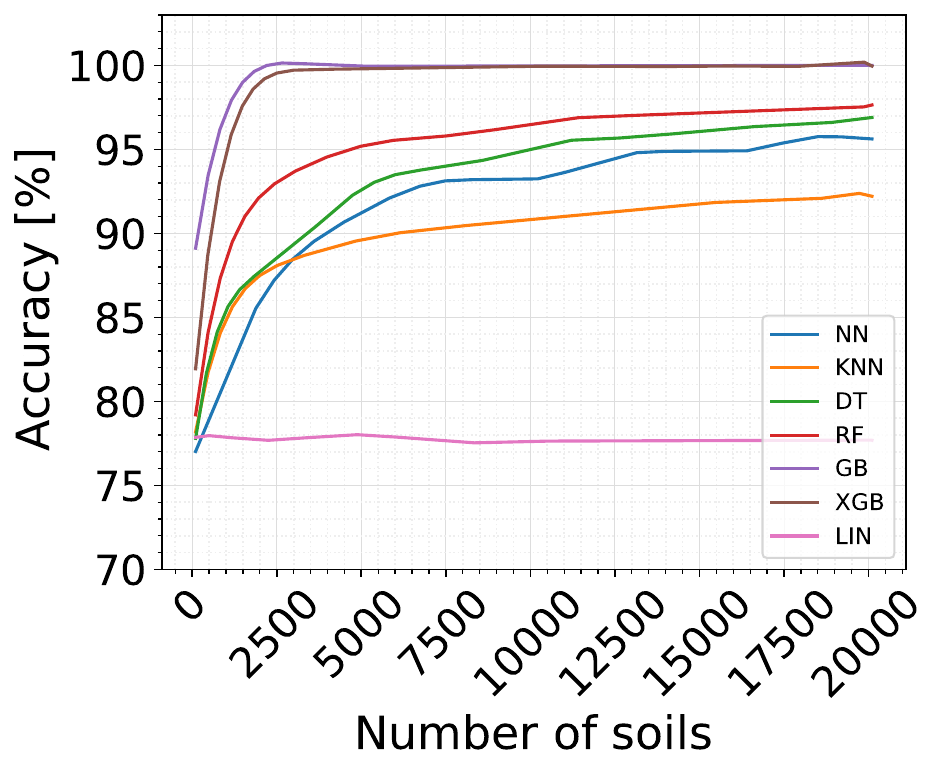}
 \caption{Application prediction performance vs number of soils.}
\label{fig:accvssoils}
\vspace{-0.3 cm}
\end{figure}

We now analyze the results for the plant recommendation application, in terms of accuracy, mean squared error (MSE), the time required for model training and inference time. The same dataset was used for both collaborative filtering and the present analysis. The performance of each algorithm was assessed using data sets containing an increasing number of soil samples, with a starting sample size of 100 and a final sample size of 22000, with increments of 1000. Figure \ref{fig:accvssoils} analyzes the accuracy of the seven algorithms vs. the number of soil samples. The datasets employed were trained on a reference set of ratings representing the optimal values. The results demonstrate that all tested algorithms exhibit a high degree of accuracy, with a range of 77\% to 99 \% on this dataset. The two algorithms, GB and XGB, achieve an accuracy of 99\% after approximately 2500 soil samples, while the linear algorithm remains at a constant value of about 77\% for all tested soil samples. The remaining algorithms exhibit a comparable pattern, demonstrating a slight increase in accuracy with an augmented soil count. However, all algorithms attain values exceeding 90 \% after 7500 soil samples.

\par
\begin{figure}
 \centering 
   \includegraphics[scale=0.4]{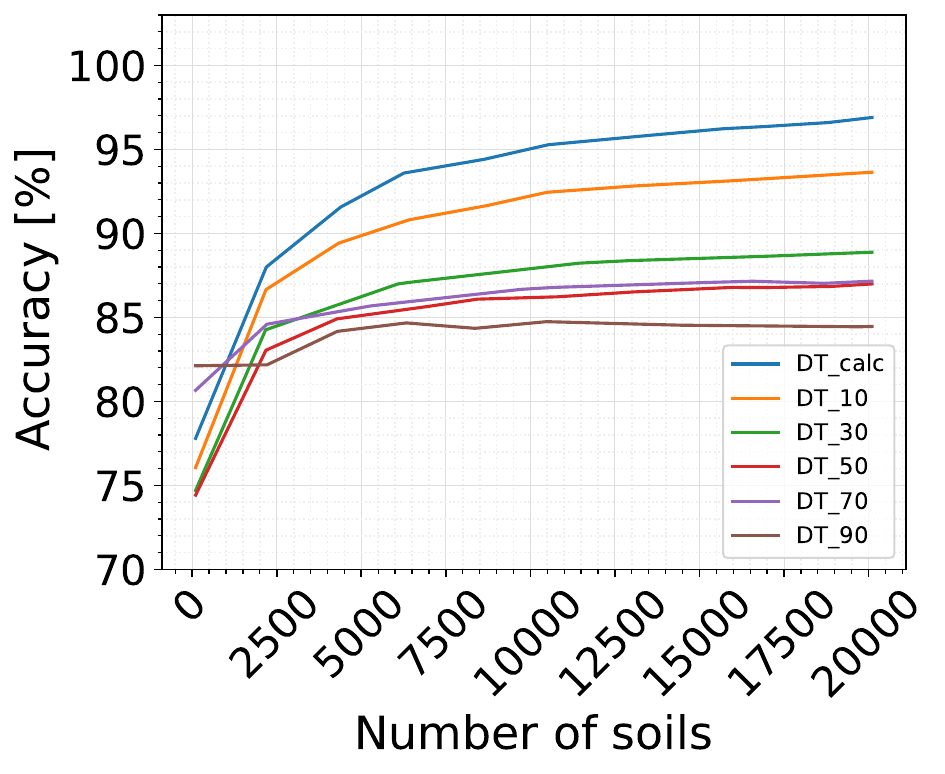}
 \caption{Applicaiotn prediction performance vs. number of soils using DT with different application data sparsity levels.}
\label{fig:DTsoils}
\vspace{-0.3 cm}
\end{figure}

 Table \ref{tab:algo} illustrates the values for the four measured metrics resulting from the largest soil count, i.e., equal 22000. The highlighted numbers indicate the highest values for the respective metric. The results demonstrate that a high accuracy achieved is associated with the prolonged training time of several seconds. The inference time for the two high-performing algorithms, GB and XGB, is also within the two-digit millisecond range, while the remaining algorithms exhibit significantly lower inference times in the microsecond range. Random forest is an exception, with an inference time of 1.5 milliseconds. The only algorithm that requires a longer inference time than training is KNN. This discrepancy in time requirements is due to the fact that, in contrast to other algorithms, no actual learning occurs during the training phase of KNN, this only occurs during the inference phase.
An analysis of the MSE values reveals that the performance varies considerably. The majority of algorithms exhibit a value below 0.1. However, the exceptions are KNN with 0.15 and linear regression with 0.67. Notably, these two also have the lowest accuracy. In terms of the best MSE value, XGB is the superior performer with a score of 0.0008. 
Figure \ref{fig:DTsoils} shows the precision of the DT algorithm when utilizing different input data. In this context, DT\_calc represents the calculated values that correspond to the optimal ratings, whereas the labels DT\_10 to DT\_90 demonstrate the algorithms performance when presented with the datasets that are calculated using cosine similarity. As can be observed, all datasets generated by cosine similarity perform slightly less effectively that the full set of data. The dataset with a sparsity of 10 \% demonstrates the best performance, with the others ordered in a descending manner. We select DT for this illustration due to its high accuracy and comparatively minimal training and inference times, making it a candidate method for this application. 

\begin{table*}
\centering
\begin{tabular}{SSSSSSSSS} \toprule
    {\textbf{Metrics}} & {\textbf{NN}} & {\textbf{KNN}} & {\textbf{Linear Regression}} & {\textbf{Decision Tree}} & {\textbf{Random Forest}}  & {\textbf{Gradient Boost}}  & {\textbf{Extreme Gradient Boost}} \\ \midrule
 {\textbf{Accuracy [\%]}}& {95.6} & {92.2} &{77.7} & {96.8} &{97.6} & {99.92} & \textbf{{99.96}} \\\midrule
 {\textbf{MSE} } & {0.09} & {0.15} & {0.67} & {0.08} & {0.05} &  {0.003} & {\textbf{0.0008}}\\ \midrule
    {\textbf{Training time [ms]}}  & {30676.7} & \textbf{{5.6}}  &{13.3} & {135.7} &{8803.8}  & {74146.1} & {7526.9} \\ \midrule
    {\textbf{Inference time [$\mu$s]}}  & {131.3} & {216.2}  &{40.3} & \textbf{{36.7}} &{1520.3} & {11669.6} & {15841.7}  \\ \bottomrule
    
\end{tabular}
\caption{Algorithms performance.}
\label{tab:algo}
\end{table*}

\section{Conclusion}\label{sec:conclusion} 
We have presented a method to optimize LoRa transmission in edge computing with TinyML-based frequency hopping algorithm, considering channel sensing for achieving reliable communication using LoRa. We integrated this concept an IoT-edge-cloud architecture and tested its performance with a plant recommendaer application in urban computing. A mathematical model for achieving optimal frequency hopping with the objective of minimizing collisions and the number of hops required was developed, which took into account historical data regarding channel occupation, in addition to the current RSSI and SNR values.  We analyzed the complexity of the model, and showed the need for heuristic or ML-based solution. A TinyML model was developed and the most suitable location for its implementation on the end device or gateway was determined through discussion and subsequently implemeted in an experimental testbed. The performance results shown for TinyML demonstrated that the model is capable of learning patterns from the self-gathered data in a laboratory setup and of predicting with an accuracy of approximately 80 \% the most suitable channel for transmitting the subsequent LoRa message in the physical layer. The model developed has been compressed to a size that allows for deployment on an ESP32 microcontroller in conjunction with code for LoRa operation. All implementation in IoT, edge, and cloud layer with LoRa is open source.

\par Finally, we presented a case study illustrating the usefulness of the the channel hopping algorithm proposed. This case study incorporated a plant recommendation system within an urban computing continuum. The system generated recommendations regarding the suitability of a set of 100 plants for cultivation, taking into account the characteristics of individual users soils. The system was designed to accommodate data sparsity, which was introduced by individual users. This part of application processing was handled in the cloud utilizing cosine similarity, resulting in data that can be used to train machine learning models for the recommendation process in the application. The results demonstrated that with the exception of one algorithm, all tested algorithms exhibit a level of accuracy on the given dataset of 90 \% and above. 

\par In future work, we plan to test a wider range of operational applications of TinyML devices in IoT-edge-cloud continuum with LoRa, towards a fully open source system open for experimentation. 
\bibliographystyle{IEEEtran}
\bibliography{references_2}

\end{document}